\documentclass[a4paper,10pt]{article}
\pdfoutput=1 
\usepackage{jheppub}
\usepackage[utf8]{inputenc}
\usepackage[english]{babel}
\usepackage{makeidx}
\usepackage{cleveref}
\usepackage{amsfonts}
\usepackage{enumerate}
\usepackage{mathrsfs}
\usepackage{tensor}
\usepackage{xcolor,tikz,pgfplots}
\usepackage{graphicx}

\usepackage{cancel}

\usepackage[normalem]{ulem}

\usepackage[autostyle]{csquotes}

\usepackage{youngtab}
\usepackage{caption}
\usepackage{subcaption}
\usepackage{enumitem}

\usepackage{braket}
\usepackage{amsthm}
\usepackage{todonotes}
\usepackage{hyperref}

\usetikzlibrary{matrix,calc,positioning,decorations.markings,decorations.pathmorphing,decorations.pathreplacing}
\usetikzlibrary{arrows,cd}
\usetikzlibrary{shapes.misc}
\usetikzlibrary {shapes.geometric}
\usetikzlibrary {shapes.multipart}
\usepackage{bbding}
\tikzset{
    >=stealth',
    punkt/.style={
           rectangle,
           rounded corners,
           draw=black, very thick,
           text width=7.2em,
           minimum height=2em,
           text centered},
    punkt2/.style={
           rectangle,
           rounded corners,
           draw=black!20!red, very thick,
           text width=7em,
           minimum height=2em,
           text centered},
    punktL/.style={
           rectangle,
           rounded corners,
           draw=black!20!red, very thick,
           text width=8.8em,
           minimum height=2em,
           text centered},
    pil/.style={
           ->,
           thick,
           shorten <=2pt,
           shorten >=2pt,},
    pil2/.style={
           <->,
           thick,
           shorten <=2pt,
           shorten >=2pt,},
    pil3/.style={
           -,
           thick,
           shorten <=2pt,}}

\usepackage{float}
\usepackage{booktabs}

\usepackage{bm}
\usepackage{anyfontsize}

\newcommand{\mN}{\mathcal{N}}

\newcommand{\Tr}{\text{Tr}}

\renewcommand{\vec}{\overrightarrow}

\newcommand{\C}{\mathbb{C}}
\newcommand{\Z}{\mathbb{Z}}

\todostyle{classic}{color=white,textcolor=blue, bordercolor=white}

\numberwithin{equation}{section}

\title{Inherited non-invertible duality symmetries\\ in quiver SCFTs}

\author[a]{Riccardo Argurio,}
\author[a]{Andrés Collinucci,}
\author[a]{Salvo Mancani,} 
\author[b,c]{Shani Meynet,}
\author[a]{Louan Mol,}
\author[d]{\\ Valdo Tatitscheff~}

\affiliation[a]{Physique Théorique et Mathématique and International Solvay Institutes,\\
Université Libre de Bruxelles, C.P. 231, 1050 Brussels, Belgium}
\affiliation[b]{Department of Mathematics, Uppsala University\\
Box 480, SE-75106 Uppsala, Sweden}
\affiliation[c]{Centre for Geometry and Physics, Uppsala University \\
Box 480, SE-75106 Uppsala, Sweden}
\affiliation[d]{Institute for Mathematics,
Ruprecht-Karls-Universität Heidelberg,\\
Mathematikon, Im Neuenheimer Feld 205, 69120 Heidelberg, Germany}

\emailAdd{riccardo.argurio@ulb.be, collinucci.phys@gmail.com, mancanisalvo@gmail.com, shani.meynet@math.uu.se, louan.mol@ulb.be, valdo.tatitscheff@normalesup.org}

\abstract{We revisit the construction of the duality group for $\mathcal N=2$ $\widehat{A}_n$-shaped quivers SCFTs and generalize it to the previously unexplored case of $\widehat{D}_n$-shaped quivers. We then provide a systematic description of non-invertible duality symmetries in both classes. Furthermore, we characterize the $\mathcal N=1$ mass deformations of these theories that preserve such symmetries, thereby identifying a large class of 
$\mathcal N=1$ SCFTs with non-invertible duality symmetries inherited from their parent 
$\mathcal N=2$ theories.} 

\preprint{}
 \clearpage

\begin{document}

\maketitle
\clearpage

\section{Introduction}

In recent years, symmetries in the context of Quantum Field Theory (QFT) have received a new paradigmatic formulation as topological defects \cite{Gaiotto:2014kfa}. One of the most noticeable consequences of this paradigm is the existence of \emph{non-invertible symmetries}. Such symmetries had already been described in the context of $d=2$ Rational Conformal Field Theories (RCFTs) \cite{Verlinde:1988sn,Moore:1988qv,Frohlich:2004ef,Frohlich:2006ch} and the $d=3$ Topological Quantum Field Theories (TQFTs) related to them \cite{Elitzur:1989nr,Fuchs:2002cm,Fuchs:2012dt}. Instead of having a group like structure, non-invertible symmetries enjoy a ring-like one, $a \times b = \sum_i c_i$, and therefore not all elements admit an inverse. Topological operators enjoying such fusion rules, despite not being invertible symmetries, can still be used to study RG-flows and Ward identities of QFTs, putting constraints on  the dynamics of the system \cite{Chang:2018iay}.

In $d=4$, a particularly helpful playground to study these generalized symmetries is $\mathcal{N}=4$ super Yang-Mills (SYM), which does in fact admit non-invertible symmetries. More precisely, $\mathcal{N}=4$ SYM enjoys a duality group, $\mathrm{SL}(2,\mathbb{Z})$, that acts on the complexified gauge coupling $\tau_{\text{SYM}}$ via modular transformations. This is the so-called Montonen--Olive duality \cite{Montonen:1977sn,Kapustin:2006pk}. Remarkably, this $\mathrm{SL}(2,\mathbb{Z})$ action admits special values of $\tau_{\text{SYM}}$, namely $i$ and $e^{\frac{2\pi i}{3}}$, that are left invariant under a discrete subgroup: $\mathbb{Z}_2$ and $\mathbb{Z}_3$ respectively. At those fixed points of the conformal manifold, duality transformations become non-invertible symmetries of the theory \cite{Kaidi:2021xfk,Choi:2021kmx, Choi:2022zal}. The non-invertibility stems from the fact that the symmetry is actually the composition of a duality transformation with a topological manipulation that reverses the effect of the self-duality transformation on the gauge group, i.e.~it relates the two different global variants \cite{Aharony:2013hda} of the gauge algebra by gauging a 1-form symmetry. The literature on the subject is vast, for a sample see \cite{Bhardwaj:2017xup, Thorngren:2019iar, Komargodski:2020mxz, Nguyen:2021naa, Thorngren:2021yso, Huang:2021zvu, Benini:2022hzx, Roumpedakis:2022aik, Bhardwaj:2022yxj, Hayashi:2022fkw, Kaidi:2022uux, Choi:2022jqy, Cordova:2022ieu, Damia:2022seq, Damia:2022bcd, Choi:2022rfe, Lin:2022dhv, Apruzzi:2022rei, Kaidi:2022cpf, Niro:2022ctq, Antinucci:2022vyk, Kaidi:2023maf, Amariti:2023hev, Copetti:2023mcq, Cordova:2023bja, Antinucci:2023ezl, Bhardwaj:2023bbf, Damia:2024xju, Heckman:2024obe, DelZotto:2024tae, Okada:2024qmk, Franco:2024mxa, Arbalestrier:2024oqg, Gutperle:2024vyp, Hasan:2024aow, Bharadwaj:2024gpj}. 

A richer set of theories, constrained enough to be reliably studied, are the so-called class $\mathcal{S}$ theories of \cite{Gaiotto:2009we}.  
The data describing these theories is encoded in Riemann surfaces with marked points and they admit, as $\mathcal{N}=4$ SYM, a duality group given by the Mapping Class Group (MCG) of the surface. As one might expect, these theories also admit non-invertible symmetries precisely when the duality group preserves the couplings of the theory while altering the global structure of the gauge group, see for instance \cite{Bashmakov:2022jtl, Bashmakov:2022uek, Antinucci:2022cdi, Carta:2023bqn} for a study of some class $\cal S$ theories. 

Symmetries, including non-invertible ones, are particularly interesting if they are preserved along an RG flow, since they can then constrain, or predict, some properties of the IR theory at the end of the flow. The simplest RG flows are those triggered by mass terms. In $\mN=2$ (and $\mN=4$) theories, the latter usually partially break supersymmetry to $\mN=1$. If the $\mN=2$ theory enjoys non-invertible symmetries, one can turn on mass terms that preserve them, and the IR theory is then expected to enjoy the same non-invertible symmetries. The analysis depends on whether the IR theory is gapped, or an SCFT. A class of gapped RG flows was considered in \cite{Damia:2023ses}. The case of the flows from some specific $\mN=2$ class $\cal S$ theories to $\mN=1$ SCFTs was also briefly considered in the same reference, see also \cite{Cordova:2023her, Cordova:2024vys, Antinucci:2024ltv, DelZotto:2024arv}.

One of the aims of this paper is to systematically discuss duality symmetries, and the mass deformations preserving them, in two broad classes of $\mN=2$ SCFTs, namely the $\widehat{A}_{n-1}$ and the $\widehat{D}_n$ quiver gauge theories. Such class $\cal S$ theories have appeared in string/M-theory constructions: they can be alternatively seen to arise from D3-branes at $\mathbb{C}^2/\Gamma\times\mathbb{C}$ singularities in type IIB \cite{Douglas:1996sw}, from D4-branes suspended between NS5-branes in type IIA \cite{Witten:1997sc}, and from M5-branes wrapping complex surfaces in M-theory \cite{Witten:1997sc}. All such descriptions are related to each other by string dualities.

Starting from the $\widehat{A}_{n-1}$ quiver gauge theory, in \cref{sec:dualitygroups} we describe the duality group in terms of the mapping class group of a complex torus with $n$ unordered marked points, revisiting the analysis of \cite{Halmagyi:2004ju}. This is best understood from the M-theory uplift. Using this result, it is possible to classify point configurations that are invariant under the duality group, leading to non-invertible symmetry defects \cite{Damia:2023ses}. We then generalize this construction to the $\widehat{D}_n$ quiver gauge theories, which has been mostly discussed in the type IIA setting \cite{Kapustin:1998fa,Hanany:2000fq,Chacaltana:2012ch}. The M-theory uplift of this theory consists of M5-branes inserted as marked points on a quotiented torus, i.e.~a {\em pillowcase}. The duality group is the mapping class group of this object, which we determine. We then proceed to study the presence of non-invertible defects in these models as well, again by finding which elements of the MCG fix the modular parameter and the marked points. Finally, despite not having a Riemann surface describing the $\widehat{E}_n$ quiver theories,\footnote{See \cite{Carta:2022spy} for an attempt towards this goal.} what we learned from the other cases allows us to make an educated guess of the structure for their duality groups. 

In \cref{sec:MassDefN1}, we turn our attention to mass deformations of $\widehat{A}_{n-1}$ and $\widehat{D}_n$ quiver gauge theories, studying the action of the duality group on them. In both cases, an important distinction is made whether an overall mass parameter, the “global mass", is zero or not. In the former case, only permutations of points act on the masses, while in the latter, $\mathrm{SL}(2,\mathbb{Z})$ transformations act on them as well. This overall mass parameter plays an important role also in determining the moduli space of the SCFT that is supposed to exist in the IR of such RG flows, as we discuss in \cref{sec:modulispace}. Indeed, the moduli space of the starting theory is a three-fold algebraic variety given by the direct product of a \emph{Du Val} type singular surface times $\mathbb{C}$. After the mass deformation, the flow brings the theory to another supersymmetric theory, whose moduli space is either a \emph{compound Du Val} three-fold or a \emph{Du Val} two-fold. The former (locally) is a non-trivial fibration of $\mathbb{C}^2/\Gamma$ over $\mathbb{C}$, while the latter is just $\mathbb{C}^2/\Gamma$. A vanishing global mass triggers a deformation leading to three-dimensional moduli space, while a non-vanishing one leads to a two-dimensional one.

In \cref{sec:dualitydefects}, we finally turn our attention to mass deformations that preserve duality defects. We prove that such a deformation always exists for both the  $\widehat{A}_{n-1}$ and $\widehat{D}_n$. We find all the non-invertible defect-preserving mass deformations for those theories, discussing some selected examples. Our main result is thus a characterization of ${\cal N}=1$ SCFTs which enjoy non-invertible duality symmetries inherited from their ${\cal N}=2$ parent SCFTs.

\section{\texorpdfstring{$\mathcal N=2$}{N=2} quiver SCFTs, dualities and symmetries}\label{sec:dualitygroups}

In this section, we first review the brane/geometric construction that at low energy leads to an affine 4d $\mathcal N=2$ ADE-type quiver SCFTs, and then also discuss the form of their duality groups. These theories are well known in the context of the AdS/CFT correspondence, where they are realized as the world volume theory of D3-branes probing Du Val surfaces, i.e.~orbifolds of the form $\mathbb{C}^2/\Gamma$ with $\Gamma$ a finite subgroup of $\mathrm{SU}(2)$. Moreover, at least $\widehat{A}_{n-1}$ and $\widehat{D}_n$ quivers also admit a class $\mathcal S$ realization which is crucial for understanding their duality groups, as we will review shortly.

In the framework of theories of class $\mathcal S$, 4d $\mathcal{N}=2$ theories are engineered by wrapping M5-branes on a genus $g$ Riemann surface with $n$ marked points, to which we will also refer as punctures, with a partial topological twist \cite{Gaiotto:2009we}. Let $\Sigma_{g,n}$ be the underlying smooth surface.\footnote{A smooth surface is a smooth manifold of real dimension two. We only consider surfaces of finite type. Smooth surfaces of finite type are entirely determined by their genus $g$ and number of punctures $n$. A Riemann surface is a smooth surface endowed with a complex structure. In general there are many inequivalent complex structures with which a fixed smooth punctured surface can be endowed; more precisely, the real dimension of the Teichmüller space $\mathcal T(\Sigma_{g,n})$ is $6g-6+2n$. Given a Riemann surface in $\mathcal T(\Sigma_{g,n})$, we interpret the punctures as marked points.} When the construction leads to an SCFT, the conformal manifold of the latter is the Teichmüller space $\mathcal{T}(\Sigma_{g,n})$. By definition, $\mathcal{T}(\Sigma_{g,n})$ is the space of all complex structures with which $\Sigma_{g,n}$ can be endowed, up to diffeomorphims of $\Sigma_{g,n}$ homotopic to the identity. The duality group of the theory is then embodied as the mapping class group $\mathrm{MCG}(\Sigma_{g,n})$, which is the group of orientation-preserving diffeomorphims of $\Sigma_{g,n}$, modulo diffeomorphisms connected to the identity.\footnote{This can be described as the group of “large" diffeomorphisms modulo “small" ones, borrowing the usual gauge theory nomenclature for transformations that cannot or can, respectively, be deformed to the identity.} Indeed, the MCG does not change the physical properties of the configuration, but it acts non-trivially on $\mathcal{T}(\Sigma_{g,n})$, relating different points of the conformal manifold \cite{Gaiotto:2009we}. We refer to \cite{Akhond:2021xio} for a general introduction to class $\mathcal S$ theories.

After computing the duality groups of $\widehat{A}_{n-1}$ and $\widehat{D}_n$ quivers, and proposing a description for $\widehat{E}_{6,7,8}$ quivers, we discuss the interplay between dualities and 1-form symmetries, and compute the locus in the conformal manifold at which non-invertible symmetries are realized.

\subsection{\texorpdfstring{$\widehat{A}_{n-1}$}{A(n-1)} quivers from class \texorpdfstring{$\mathcal{S}$}{S}}\label{sec:N2An}

\begin{figure}[h!]
    \centering
    \includegraphics[scale=0.8]{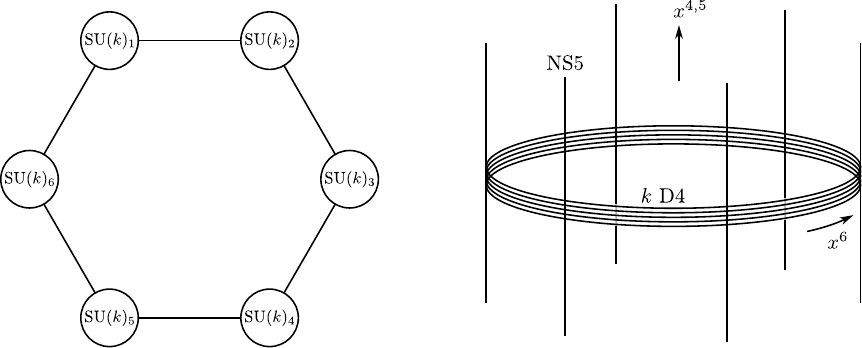}
    \caption{“Necklace" $\mathcal N=2$ quiver and corresponding type IIA brane configuration.}
    \label{fig:typeAquiver}
\end{figure}

We first consider $4d$ $\mathcal N=2$ quivers gauge theories shaped like affine $\widehat{A}_{n-1}$ Dynkin diagrams, which we will refer to as $\widehat{A}_{n-1}$ theories for convenience. These consist of $n$ $\mathrm{SU}(k)$ gauge factors with bifundamental hypermultiplets, as shown in \cref{fig:typeAquiver}. We denote the gauge factors as $\mathrm{SU}(k)_1, \dots, \mathrm{SU}(k)_{n}$, and for each $i=1,\dots,n$ modulo $n$ there is an adjoint field $\phi_i$ and a pair of chiral multiplets $X_{i,i+1}$, $X_{i+1,i}$ in the representation $\left( \tiny{\yng(1)}_{i}, \overline{\tiny{\yng(1)}}_{i+1} \right)$, $\left( \tiny{\yng(1)}_{i+1}, \overline{\tiny{\yng(1)}}_{i} \right)$ of $\mathrm{SU}(k)_{i}\times \mathrm{SU}(k)_{i+1}$. The superpotential is the minimal one compatible with $\mathcal{N}=2$ supersymmetry, that is:
\begin{align}
    W_{\mathcal{N}=2} = \sum_{i=1}^{n} \phi_i \left( X_{i,i+1}X_{i+1,i} - X_{i,i-1}X_{i-1,i} \right) \; .
\end{align}

These theories can be realized in type IIA string theory as the worldvolume theory of a stack of $k$ D4-branes suspended between $n$ NS5-branes along a circle: these are the elliptic models of \cite[Section 4]{Witten:1997sc}. More precisely, one considers type IIA string theory in $\mathbb{R}^{1,3}\times\mathbb{R}^2_{4,5}\times S^1_6\times\mathbb{R}^3_{7,8,9}$, with $n$ NS5-branes extending along $\mathbb{R}^{1,3}\times\mathbb{R}^2_{4,5}$ and $k$ D4-branes along $\mathbb{R}^{1,3}\times S^1_6$, all at the same point in $\mathbb{R}^3_{7,8,9}$, as shown in \cref{tab:braneelliptic}.

\begin{table}[h!]
    \centering
    \begin{tabular}{c|cccccccccc}
         & 0 & 1 & 2 & 3 & 4 & 5 & 6 & 7 & 8 & 9 \\ \hline
       $n$ NS5 & $-$ & $-$ & $-$ & $-$ & $-$ & $-$ & & $\cdot$ & $\cdot$ & $\cdot$ \\
       $k$ D4 & $-$ & $-$ & $-$ & $-$ & & & $-$ & $\cdot$ & $\cdot$ &  $\cdot$
    \end{tabular}
    \caption{Type IIA brane configurations defining elliptic models.}\label{tab:braneelliptic}
\end{table}

A cartoon of this brane setup in  $\mathbb{R}^2_{4,5}\times S^1_6$ is shown on the right hand side of \Cref{fig:typeAquiver}. 
A dual description in type IIB string theory of this configuration is obtained as the worldvolume theory of a stack of $k$ D3-branes transverse to $\mathbb{C}^2/\mathbb{Z}_{n} \times \mathbb{C}$. The two descriptions are related by T-duality along $x^6$ and decompactification.

Such theories are superconformal at any value of the gauge couplings. The inverse gauge coupling squared of a given node is proportional to the distance between the corresponding NS5-branes along $x^6$ \cite{Witten:1997sc}. More precisely, if the circle $x^6$ has length $2\pi R_6$ and $g_s$ denotes the string coupling, then:
\begin{equation}
    \frac{1}{g_i^2} = \frac{x_{i+1}^6-x_i^6}{8\pi g_s R_6}~.
\end{equation}
We are now interested in the uplift of such configurations to M-theory, where the relationship between the elliptic brane model and the class $\mathcal{S}$ construction is manifest.

Let $2\pi R_{10}$ be the length of the M-theory circle $S^1_{10}$. As emphasized in \cite{Witten:1997sc}, the metric on $S^1_6\times S^1_{10}$ is not necessarily the product metric: the shift $x^6\rightarrow x^6+2\pi R_6$ can in general be accompanied by a shift $x^{10}\rightarrow x^{10}+\theta R_{10}$, where $\theta$ is some angle. Let:
\begin{equation}
    \tau = \frac{\theta}{2\pi} + \frac{i}{8\pi g_s}~,\label{eq:couplexcoupl}
\end{equation}
so that the torus metric is the natural flat metric on the elliptic curve $E_\tau$ with modulus $\tau$ in the complex upper-half plane $\mathbb{H}$. 

\begin{figure}[h!]
    \centering
    \includegraphics[width=\textwidth]{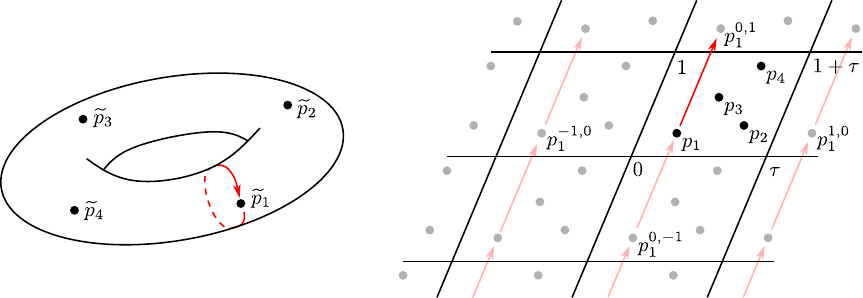}
    \caption{Riemann surface corresponding to the $\widehat{A}_3$ quiver theory and its universal cover.}
    \label{fig:typeAsetup}
\end{figure}

In the uplift to M-theory both D4 and NS5-branes become M5-branes: the former correspond to M5-branes wrapping the elliptic curve $E_\tau$, whereas the latter are interpreted as boundary conditions for the worldvolume theory on the stack of $k$ M5-branes at marked points on $E_\tau$. Thus, this set-up can be reformulated in the class $\mathcal{S}$ framework, where the Riemann surface is the elliptic curve $E_\tau$ with $n$ marked points, with underlying smooth surface  $\Sigma_{1,n}$. Let $\widetilde p_1,\dots \widetilde p_n \in E_\tau$ denote the positions of the marked points as in \cref{fig:typeAsetup}.

The mapping class group of $\Sigma_{1,n}$ can be obtained from the one of the closed torus $\mathbb{T}^2$ via the Birman exact sequence \cite{Birman:1969mcg}, and it can be described explicitly as follows. We consider the universal cover of the curve $E_\tau$ together with the lifts of the marked points, as depicted on the right of \cref{fig:typeAsetup}. The elliptic curve $E_\tau$ can be presented as $E_\tau=\mathbb{C}/\Lambda$, where $\Lambda\subset\mathbb{C}$ is the lattice $\mathbb{Z}+\tau\mathbb{Z}$. Let us fix a starting notation for the lifts of $\widetilde p_1,\dots \widetilde p_n \in E_\tau$: $p_1:=p_1^{0,0},\dots,p_n:=p_n^{0,0}$ denote the lifts in the fundamental parallelogram $\{0,1,\tau,1+\tau\}$, while those in the parallelogram $\{k+l\tau,k+1+l\tau,k+(l+1)\tau,k+1+(l+1)\tau\}$ are denoted $p_1^{k,l},\dots,p_n^{k,l}$. When the configuration of marked points on $E_\tau$ is generic,\footnote{By generic, we refer to configurations of marked points that do not satisfy any special constraints. In particular, we assume that the line passing through any two arbitrary points in the configuration is never parallel with any line connecting two points of the lattice $\Lambda$. This guarantees that, in any duality frame, the points can be labeled as $p_1,\dots,p_n$, in such a way that the condition $\mathrm{Im}(p_1)<\mathrm{Im}(p_2)<\dots<\mathrm{Im}(p_n)$ holds. Non-generic configurations form a measure-zero subset of the space of all possible configurations, justifying the terminology.} there is a way to label them that is suitable for the physical interpretation of the setup, which is such that $\mathrm{Im}(p_1)<\mathrm{Im}(p_2)<\dots<\mathrm{Im}(p_n)$. We will explain in which sense it is suitable for physics shortly.

\begin{figure}[h!]
    \centering
    \includegraphics[]{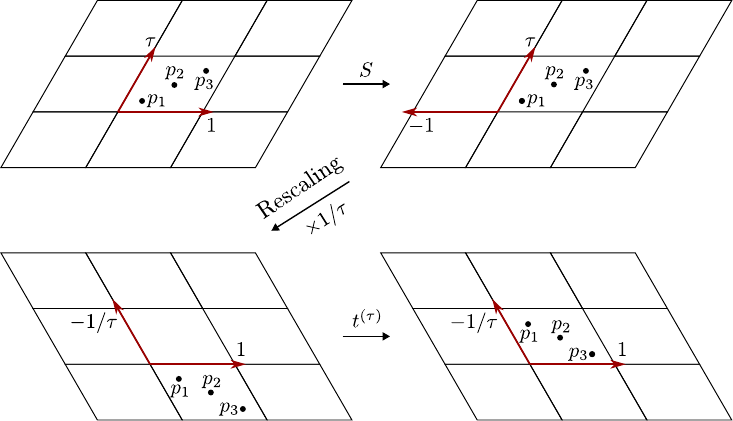}
    \caption{We start from the top left configuration. We first change the basis of the lattice $\Lambda$, but not the position of the punctures, then rescale the vector space $\mathbb{C}$ and finally shift the punctures back to the fundamental cell. For the sake of clarity, we omitted all the lifts of the punctures, but the ``fundamental" ones.}\label{fig:a2example}
\end{figure}

The generators of the mapping class group are of three types:

\begin{enumerate}[label=\textit{\arabic*)}]
    \item \textbf{Mapping classes of the torus.}\quad The modular group $\mathrm{SL}(2,\mathbb{Z})$ acts as a change of basis for the lattice $\Lambda$. The action of the standard generators $T$ and $S$ is the following: $T:(1,\tau)\rightarrow (1,\tau+1)$, whereas $S$ encodes the combined operation $(1,\tau)\rightarrow (\tau,-1)\simeq (1,-1/\tau)$. Because of the rescaling by $1/\tau$, the generator $S$ acts non-trivially on the marked points: if $p\in \mathbb{C}$ is a lift of a puncture then $S \, p = p/\tau$. The action of $S$ is depicted in \cref{fig:a2example}.
    
    \item \textbf{Deck transformations.}\quad They are defined as changing the choice of lifts of the marked points,\footnote{Here we use the standard expression “deck transformations" from the theory of coverings in a loose sense, as actual deck transformations would act on all marked points together. However, this terminology makes manifest the relation between the fundamental group of the Riemann surface and its braiding action on a given marked point.} and are generated by $t_i^{(1)}:p_i\rightarrow p_i+1$ and $t_i^{(\tau)}:p_i\rightarrow p_i+\tau$ for $i=1,\dots,n$. The red arrows in \cref{fig:typeAsetup} depict the action of $t_1^{(\tau)}$. In other words after acting with a deck transformation the lifts denoted $p_i$, $i=1,\dots,n$, are not necessarily in the fundamental parallelogram $\{0,1,\tau,1+\tau\}$ anymore.
                
    \item \textbf{Permutations of the punctures.}\quad We describe these transformations in the universal cover of the torus; the generators are denoted $s_i$, $i=1,\dots,n$, where $s_i$ for any $i=1,\dots,n-1$ exchanges $p_i$ and $p_{i+1}$, whereas $s_n$ exchanges $p_n$ with $p_1+\tau = p_1^{0,1}$. This specificity in the definition of $s_n$ echoes in \cref{eq:mcgpoint} below.
\end{enumerate}

Denoting $\vec{p} = (\tau ;  p_1, \dots, p_n)$, the generators of the mapping class group act as 
\begin{align}\label{eq:mcgpoint}
    S \; : \; \vec{p} &\longmapsto \left( -\frac{1}{\tau}; \frac{p_1}{\tau}, \frac{p_2}{\tau}, \ldots, \frac{p_n}{\tau} \right) \; , \nonumber \\[5pt]
    T \; : \; \vec{p} &\longmapsto \left( \tau + 1; p_1, p_2, \ldots, p_n \right) \; , \nonumber \\[5pt]
    t_i^{(1)} \; : \; \vec{p} &\longmapsto \left( \tau; p_1, p_2, \ldots, p_i + 1, \ldots, p_n \right) \; , \nonumber \\[5pt]
    t_i^{(\tau)} \; : \; \vec{p} &\longmapsto \left( \tau; p_1, p_2, \ldots, p_i + \tau, \ldots, p_n \right) \; , \nonumber \\[5pt]
    s_i \; : \; \vec{p} &\longmapsto \left( \tau; p_1, p_2, \ldots, p_{i-1}, p_{i+1}, p_i, p_{i+2},\ldots, p_n \right) \; , \nonumber \\[5pt]
    s_n \; : \; \vec{p} &\longmapsto \left( \tau; p_n - \tau , p_2, \ldots, p_1 +\tau \right) \; ,
\end{align}

This transposes to the physical theory as follows. After lifting to M-theory, the complexified gauge couplings of each node of the quiver are recovered as differences in position of neighboring marked points on $E_\tau$ \cite{Witten:1997sc}. Let
\begin{align}\label{eq:deftaui}
\begin{cases}
    \tau_i = p_{i+1}-p_{i} \, , \, i \neq n \\
    \tau_n = \tau + p_1 - p_n \,
\end{cases}
\end{align}
where the seemingly special definition of $\tau_n$ follows from the fact that $\sum \tau_i = \tau$. The ``physical" labeling of the punctures described above ensures that $4\pi\mathrm{Im}(\tau_i) = g_i^{-2} >0$ for all $i$, where $g_i$ is the gauge coupling of the $i$-th gauge group of the quiver gauge theory.\footnote{As discussed in \cite{Damia:2023ses} (see also \cite{Halmagyi:2004ju}), for a given starting choice of punctures, only a subgroup of the MCG can be rightfully labeled as the duality group, namely the combinations of the above operations that preserve the imaginary ordering of the punctures. We prefer to describe the full MCG, which does not depend on the initial choice of punctures.}

Given \cref{eq:deftaui} and denoting $\vec{\tau}=(\tau;\tau_1,\dots,\tau_{n})$, one can recast the action of the generators of the MCG as
\begin{align}\label{eq:mcgtau}
    &S \; \; \; \, : \; \vec{\tau} \to \left( -\frac{1}{\tau}; \frac{\tau_1}{\tau}, \frac{\tau_2}{\tau}, \ldots, \frac{\tau_n}{\tau} - 1 - \frac{1}{\tau} \right) \; , \nonumber \\[5pt]
    &T \; \; \; \, : \; \vec{\tau} \to \left( \tau + 1; \tau_1, \tau_2, \ldots, \tau_n + 1 \right) \; , \nonumber \\[5pt]
    &t_i^{(1)} \; : \; \vec{\tau} \to \left( \tau; \tau_1, \tau_2, \ldots, \tau_{i-1} + 1, \tau_i - 1, \ldots \tau_n \right) \; , \nonumber \\[5pt]
    &t_i^{(\tau)} \; : \; \vec{\tau} \to \left( \tau; \tau_1, \tau_2, \ldots, \tau_{i-1} + \tau, \tau_i - \tau, \ldots, \tau_n \right) \; , \nonumber \\[5pt]
    &s_i \;\;\; \, : \; \vec{\tau} \to \left( \tau; \tau_1, \tau_2, \ldots, \tau_{i-1} + \tau_i, -\tau_i, \tau_{i+1} + \tau_i, \ldots, \tau_n \right) \; .
\end{align}
All $s_i$ together generate the affine Weyl group of type $\widehat{A}_{n-1}$, as is usual for brane configurations on a circle \cite{Hanany:2001iy}. Together with the transformations $t_i^{(\tau)}$, they generate the group of automorphisms of the $\widehat{A}_{n-1}$ root system, that is, the co-central extension of the affine Weyl group of type $\widehat{A}_{n-1}$ by the group of outer automorphisms of the affine Lie algebra of type $\widehat{A}_{n-1}$. We refer to \cite{Kac:1990gs,DiFrancesco:1997nk} for more details on affine Lie algebras and Weyl groups, here we simply discuss how the correspondence is achieved. Let us consider the vertical band of fundamental parallelograms containing the vertices $\{0,1,\tau,1+\tau\}$. The differences $p_i^{0,a}-p_j^{0,b}$ where $i,j=1,\dots,n$ and $a,b\in\mathbb{Z}$, define the affine root lattice of type $\widehat{A}_{n-1}$. The standard positive simple roots are the $\tau_i$ defined in \cref{eq:deftaui}, where $\tau_n = \tau_0$ is the affine simple root and the shift by $\tau$ embodies the single imaginary root of the affine root system. The whole group of automorphisms of the lattice is generated by the $s_i$ and $t_i^{(\tau)}$, for $i=1,\dots,n$.\footnote{Let us note that there are relations between the generators of the MCG, for example $t_i^{1}=S^{-1}(t_i^{\tau})^{-1}S$. Moreover, the outer automorphism $\omega$ of the $\widehat{A}_{n-1}$ algebra can be expressed as $\omega = t_1^{(\tau)}s_1s_2\dots s_{n-1}$.}

The description of the MCG in terms of the automorphisms of an affine root system will be used in \cref{sec:enduality} to argue the structure of the duality group for theories for which an explicit class $\mathcal{S}$ construction is not known. 

\subsection{\texorpdfstring{$\widehat{D}_n$}{Dn} quivers from class \texorpdfstring{$\mathcal{S}$}{S}}\label{sec:DnUplifts}

\begin{figure}[h!]
    \centering
    \includegraphics[]{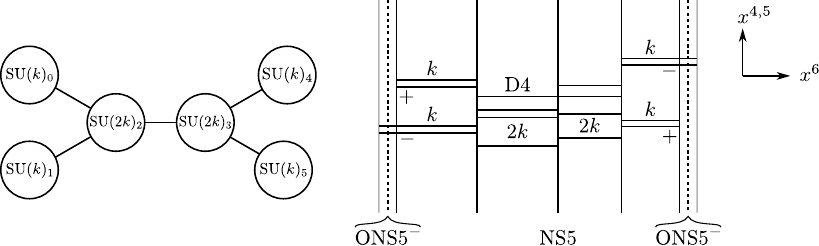}
    \caption{$\widehat{D}_5$ quiver and corresponding type IIA brane configuration.}
    \label{fig:Dnconfig}
\end{figure}

We now turn to the description of the duality group of $\widehat{D}_n$ quiver gauge theories. An example of such a quiver is shown on the left of \cref{fig:Dnconfig}. The gauge group of the theory is a product of $(n-3)$ copies of $\mathrm{SU}(2k)$ and $4$ copies of $\mathrm{SU}(k)$, there is an adjoint field $\phi_i$ for each gauge factor and there are bifundamental hypermultiplets $X_{i,j}$ as to make the corresponding affine quiver of type $\widehat{D}_n$, and the superpotential required by $\mathcal N=2$ supersymmetry
\begin{align}
   W_{\mathcal{N}=2} &= \sum_{i=0,1} \phi_i X_{i,2} X_{2,i} + \sum_{j=n-1,n} \phi_{j} X_{j,n-2} X_{n-2,j} + \phi_2 \left( X_{2,0} X_{0,2} + X_{2,1} X_{1,2} + X_{23} X_{32} \right) \nonumber \\
    & + \phi_{n-2} \left( X_{n-2,n-1} X_{n-1,n-2} + X_{n-2,n} X_{n,n-2} - X_{n-2,n-3}X_{n-3,n-2} \right) \, \nonumber \\
    & + \sum_{l=3}^{n-3} \phi_l \left( X_{l,l+1}X_{l+1,l} - X_{l,l-1}X_{l-1,l} \right)  \; ,
\end{align}
where we refer to \cref{fig:Dnconfig} for the index conventions of the fields.

Let us discuss the type IIA brane setups that realize these theories, following \cite{Kapustin:1998fa} (see also \cite{Chacaltana:2012ch}). The relevant configurations of branes are described in \cref{tab:IIAbranes}: $2k$ D4-branes suspended between $n$ NS5-branes along a segment with endpoints $\mathrm{ONS5}^-$-planes, these last ones being orientifold-like 5-plane magnetically charged under the Neveu--Schwarz $B_2$ (and not under the Ramond--Ramond field $C_2$).\footnote{$\mathrm{ONS5}^-$-planes are obtained from type IIA $\mathrm{O4}^-$-planes by uplift to M-theory and compactification back to type IIA string theory on a transverse orbifolded circle $S^1/\mathbb{Z}_2$ \cite{Kapustin:1998fa}.} 

\begin{table}[h!]
    \centering
    \begin{tabular}{c|cccccccccc}
         & 0 & 1 & 2 & 3 & 4 & 5 & 6 & 7 & 8 & 9 \\ \hline
       2 ONS5$^-$ & $-$ & $-$ & $-$ & $-$ & $-$ & $-$ & & $\cdot$ & $\cdot$ & $\cdot$ \\
       $n$ NS5 & $-$ & $-$ & $-$ & $-$ & $-$ & $-$ & & $\cdot$ & $\cdot$ & $\cdot$ \\
       $2k$ D4 & $-$ & $-$ & $-$ & $-$ & & & $-$ & $\cdot$ & $\cdot$ &  $\cdot$
    \end{tabular}
    \caption{Type IIA brane configuration defining a 4d gauge theories of $\mathrm{SU}(2k)$ and $\mathrm{SU}(k)$ gauge groups in the shape of a $\widehat{D}_n$ quiver. Empty cells correspond to branes having possibly distinct positions along that direction, whereas all branes are located at the same point of the $x^{7,8,9}$ space.}
    \label{tab:IIAbranes}
\end{table}

D4-branes end either on the NS5-brane closest to the ONS5$^{-}$ plane or on its image, as shown in \cref{fig:Dnconfig}. In particular, the states corresponding to D4-branes stretching between this NS5-brane and its image are projected out by the orientifold \cite{Hanany:2000fq}. This explains how one obtains the characteristic ends of the $\widehat{D}_n$ quiver. Equivalently, one can bring an NS5-brane atop each $\mathrm{ONS5}^-$-plane; the worldvolume theory on this composite object is a 6d $\mathrm{O}(2)=\mathbb{Z}_2\ltimes \mathrm{SO}(2)$ gauge theory. D4-branes ending on such a composite object carry a charge $\pm1$ for the $\mathrm{O}(2)$ gauge theory.\footnote{Another way to see this is discussed in \cite{Sen:1996na, Gaiotto:2008ak}.} The superconformal configurations are those in which the stack of $2k$ D4-branes splits in two sub-stacks of $k$ D4-branes at each half $\mathrm{ONS5}^-$-plane; equivalently half the D4-branes have charge $+$ and the other half, charge $-$. This configuration realizes the $\widehat {D}_n$ quiver theory as the worldvolume theory on the D4-branes.

Just as in the $\widehat{A}_{n-1}$ case, this quiver gauge theory can be obtained in Type IIB as the worldvolume theory of a stack of $k$ D3-branes transverse to $ \mathbb{C}^2/\Gamma_{D_n}\times\mathbb{C}$, with $\Gamma_{D_n}$ the corresponding finite subgroup of $\mathrm{SU}(2)$.

The uplift to M-theory of D4 and NS5-branes happens exactly as in the $\widehat{A}_{n-1}$ case, while the $\mathrm{ONS5}^-$ become $\mathrm{OM5}$-planes inducing the involution $I_{C_3} \mathcal{I}_5$, where $\mathcal{I}_5$ acts by reversing the coordinates transverse to the $\mathrm{OM5}$-plane and $I_{C_3}$ reverses the sign of the M-theory 3-form. The uplift yields M-theory on 
\begin{equation}
    \mathbb{R}^{1,3} \times \mathbb{R}^2 \times \left(E_\tau\times \mathbb{R}^3\right)/\mathbb{Z}_2~,
\end{equation}
where we have combined the $x_6$ direction and the M-theory circle $S^1_{10}$ into an elliptic curve $E_\tau$ as before, and with M5-branes either wrapping the torus (D4), becoming marked points (NS5) or OM5-planes located at the four fixed points on $E_\tau$ (ONS5$^-$) \cite{Kapustin:1998fa,Hanany:2000fq,Chacaltana:2012ch}.

As in the previous section, this construction allows to study the duality group of the theory in term of the MCG of the quotiented torus. To this end we now turn to the description of the quotient geometry, which will give us insight on how to construct the MCG.

\subsubsection{Quotient surface and its universal cover}\label{sec:pillow}
        
\begin{figure}[h!]
    \centering
    \includegraphics[scale=1]{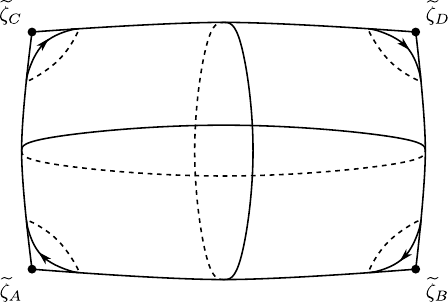}
    \caption{The pillowcase $\mathbb{T}^2/\mathbb{Z}_2$.}\label{fig:pillowcase}
\end{figure}

The action $z\mapsto-z$ on the elliptic curve $\mathbb{C}/\Lambda$, where $\Lambda = \mathbb{Z}+\tau\mathbb{Z}$, has four fixed points in the standard fundamental parallelogram $\{0,1,\tau,1+\tau\}$:
\begin{equation}
    \zeta_A = 0, \quad \zeta_B = 1/2, \quad \zeta_C = \tau/2 \quad \text{ and } \quad \zeta_D = (1+\tau)/2 \, .
    \end{equation}
The quotient space $\mathbb{T}^2/\Z_2$ is topologically a sphere with four $\mathbb{Z}_2$-orbifold points $\widetilde{\zeta}_A,\widetilde{\zeta}_B,\widetilde{\zeta}_C$ and $\widetilde{\zeta}_D$, often dubbed \emph{pillowcase} and depicted in \cref{fig:pillowcase}. Let us denote $a,b,c$ and $d$ the homotopy classes of small loops around $\widetilde{\zeta}_A,\widetilde{\zeta}_B,\widetilde{\zeta}_C$ and $\widetilde{\zeta}_D$ respectively. Since these are $\mathbb{Z}_2$-orbifold points, one has $a^2=b^2=c^2=d^2=1$. The (orbifold) fundamental group of the pillowcase is given by:
\begin{align}\label{eq:pillowp1rel}
    \pi_1 (\mathbb{T}^2/\mathbb{Z}_2) = \left\langle a,b,c,d~\left|~a^2=b^2=c^2=d^2=1, ba=dc\right.\right\rangle \, .
\end{align}
This can be obtained as follows.

\begin{figure}[h!]
    \centering
    \includegraphics[scale=1]{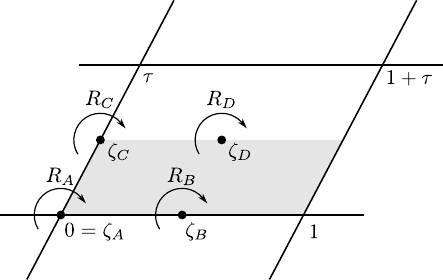}
    \caption{Generators $R_A$, $R_B$, $R_C$ and $R_D$ of $\pi_1(\mathbb{T}^2/\mathbb{Z}_2)$. A fundamental cell is shown in grey.}\label{fig:fundgrouporb}
\end{figure}

In the double cover $\mathbb{T}^2$ of the space $\mathbb{T}^2/\mathbb{Z}_2$, we can define the reflections with respect to $\zeta_A, \zeta_B, \zeta_C$ and $\zeta_D$ respectively, as depicted in \cref{fig:fundgrouporb}, which act as:
\begin{align}
    R_A &: p\mapsto -p \; ,\\
    R_B &: p\mapsto 1-p \; ,\\
    R_C &: p\mapsto \tau-p \; ,\\
    R_D &: p\mapsto (1+\tau)-p \; .
\end{align}

The fundamental group $\pi_1(\mathbb{T}^2/\mathbb{Z}_2)$ is generated by $\{R_A,R_B,R_C,R_D\}$, and one has $R_B \circ R_A (p) = R_D \circ R_C (p) = p + 1 = t^{(1)}(p)$ and $R_C \circ R_A (p) = R_D \circ R_B (p) = p + \tau = t^{(\tau)}(p)$. 
Correspondingly, $\pi_1(\mathbb{T}^2)$ embeds in $\pi_1(\mathbb{T}^2/\mathbb{Z}_2)$ as a subgroup of order 2.

\subsubsection{The \texorpdfstring{$\widehat{D}_n$}{Dn}-ality group}

As in the $\widehat{A}_{n-1}$ case, we obtain the generators of the duality group in the $\widehat{D}_n$ case by considering the lifts of the marked points on $E_\tau/\mathbb{Z}_2$, in the universal cover $\C$. Note that $n$ marked points on $E_\tau/\mathbb{Z}_2$, none of which sits at an orbifold point, correspond to $2n$ marked points on $E_\tau$ consisting of $n$ symmetric pairs with respect to the center of the fundamental cell. 

\begin{figure}[h!]
    \centering
    \includegraphics[scale=1]{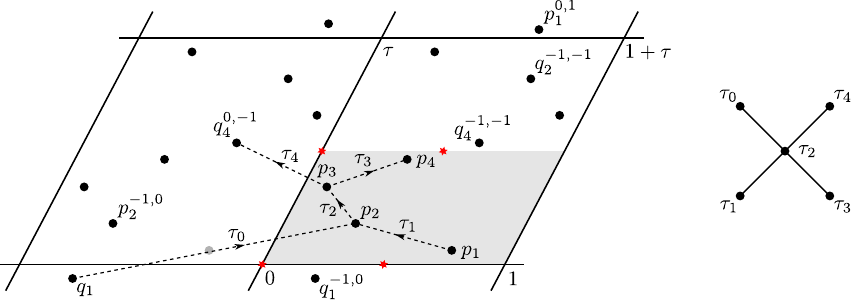}
    \caption{Setup and conventions in the specific case of the $\widehat{D}_4$ quiver.}\label{fig:Dnconfigpoints}
\end{figure}

One can choose one half of the fundamental cell, for example the bottom one as in \cref{fig:fundgrouporb}, and label the lifts of the marked points sitting in it $p_1,\dots,p_n$.\footnote{We assume the configuration of marked points to be generic.} The shifts of these lifts by elements of the lattice are as before labeled $p_i^{k,l} = p_i+k+l\tau$, with $k,l\in\mathbb{Z}$. Last, for all $i,k,l$ we let $q_i^{k,l}=-p_i^{k,l}$ be the image of $p_i^{k,l}$ reflected about the origin. There is a labeling adapted to physics, which is such that $\mathrm{Im}(p_1)<\dots<\mathrm{Im}(p_n)$. This setup is shown in \cref{fig:Dnconfigpoints}.

As before, the generators of the duality group are of three types:

\begin{enumerate}[label=\textit{\arabic*)}]
    \item \textbf{Modular group.}\quad The modular group $\mathrm{SL}(2,\mathbb{Z})$ acts as in the $\widehat{A}_{n-1}$ case:
    \begin{align}
        \mathcal S&:(\tau;p_1,\dots,p_n)\mapsto\left(-\frac{1}{\tau};\frac{p_1}{\tau},\dots,\frac{p_n}{\tau}\right) \; ,\\
        \mathcal T&:(\tau;p_1,\dots,p_n)\mapsto(\tau+1;p_1,\dots,p_n) \; .
    \end{align}
    \item \textbf{Deck transformations.}\quad The group of deck transformations of $\C\rightarrow \mathbb{T}^2/\Z_2$ is generated by $\{R_A,R_B,R_C,R_D\}$, hence each marked point is transformed as:
    \begin{align}\label{eq:DeckDpunctures}
        R_{A,i} &: (\tau;p_1,\dots,p_n)\mapsto(\tau;p_1,\dots,-p_i,\dots,p_n) \; ,\\
        R_{B,i} &: (\tau;p_1,\dots,p_n)\mapsto(\tau;p_1,\dots,1-p_i,\dots,p_n) \; ,\\
        R_{C,i} &: (\tau;p_1,\dots,p_n)\mapsto(\tau;p_1,\dots,\tau-p_i,\dots,p_n) \; ,\\
        R_{D,i} &: (\tau;p_1,\dots,p_n)\mapsto(\tau;p_1,\dots,(1+\tau)-p_i,\dots,p_n) \; ,
    \end{align}
    with the same relations as the ones that $\{R_A,R_B,R_C,R_D\}$ satisfy.
    \item \textbf{Permutation of the punctures.}\quad Punctures can be permuted, with generators:
    \begin{align}
        s_{i} &: (\tau;p_1,\dots, p_i, p_{i+1},\dots,p_n)\mapsto(\tau;p_1,\dots,p_{i+1},p_i,\dots,p_n)
    \end{align}
    for $i=1,\dots,n-1$. One could add an additional generator $s_n$ which would exchange $p_n$ and $q_n^{-1,-1}$, however $s_n=R_{D,n}$ which we have already taken into account.
\end{enumerate}
        
Similarly to what we did in the $\widehat{A}_{n-1}$ case, we now turn to the “physical" picture in which we discuss complexified gauge couplings instead of punctures, defined as follows:
\begin{align}\label{eq:couplDn}
    \tau_0 &= p_2+p_1 \; , \nonumber\\
    \tau_{i} &= p_{i+1}-p_i \; , \qquad i=1, \, \ldots \, , n-1 \nonumber\\
    \tau_n &=\tau-p_n-p_{n-1} \; .
\end{align}
Again, the physical labeling ensures that each of the $\tau_i$ has positive imaginary part. The couplings satisfy the following relation:
\begin{equation}
    \tau_0 + \tau_1 + \tau_{n-1} + \tau_n + 2 \sum_{i=2}^{n-2}\tau_i = \tau.
\end{equation}
Letting $n_0=n_1=n_{n-1}=n_n=1$ and $n_{i}=2$ for $1\leq i\leq n-1$, the previous equations rewrites as
\begin{equation}\label{eq:Dynkinlabels2}
    \sum_{i=0}^{n} n_i\tau_i = \tau \; .
\end{equation}
These weights are nothing but the Dynkin labels of the affine $\widehat{D}_n$ Dynkin graph, or equivalently, the ranks of the nodes in the McKay graph corresponding to dihedral groups.

Using the action in the marked point basis and \cref{eq:couplDn} and denoting as before $\vec{\tau}=(\tau;\tau_0,\dots,\tau_n)$, one finds that the action of the duality group in the coupling basis reads
\begin{align}
    S &: \vec{\tau} \longmapsto \left(-\frac{1}{\tau};\frac{\tau_0}{\tau},\dots,\frac{\tau_{n-1}}{\tau},\frac{\tau_n}{\tau}-1-\frac{1}{\tau}\right)~,\\
    T &:  \vec{\tau} \longmapsto  (\tau+1;\tau_0,\dots,\tau_{n-1},\tau_n+1)~,
\end{align}
and the action of a deck transformation acts as
{\small
\begin{equation}
    R_{I,i} : \vec{\tau} \longmapsto
    \begin{cases}
        (\tau; 2\zeta_I+\tau_1,\tau_0-2\zeta_I,\dots,\tau_n) & (i=1),\\[5pt]
        (\tau; 2\zeta_I-\tau_1,2\zeta_I-\tau_0,\tau_0+\tau_1+\tau_2-2\zeta_I,\dots,\tau_n) & (i=2),\\[5pt]
        \left(\tau; \tau_0,\dots,2\zeta_I-\displaystyle\sum_{k=0}^{i-1}n_k\tau_k+\tau_{i-1},-2\zeta_I-\sum_{k=0}^{i}n_k\tau_k+\tau_{i},\dots,\tau_n\right) & (3\leq i\leq n-2),\\[5pt]
        (\tau; \tau_0,\dots,2\zeta_I+\tau_{n-2}+\tau_{n-1}+\tau_n-\tau, \tau-\tau_n-2\zeta_I, \tau-\tau_{n-1}-2\zeta_I)& (i=n-1),\\[5pt]
        (\tau; \tau_0,\dots,2\zeta_I+\tau_n-\tau, -2\zeta_I+\tau_{n-1}+\tau) & (i=n),
    \end{cases}
\end{equation}}\\
for $I=A,B,C,D$. The structure of these transformations can be written more economically by using the partial sums
\begin{equation}
    P(i)=\sum_{k=0}^{i}n_k\tau_k = 2 p_{i+1}\; ,
\end{equation}
in terms of which one obtains
{\small
\begin{equation}\label{eq:deckD}
    R_{I,i} : \vec{\tau} \longmapsto
    \begin{cases}
        (\tau;2\zeta_I+P(1)-\tau_0,-2\zeta_I+P(1)-\tau_1,\tau_2,\dots,\tau_n) & (i=1),\\[5pt]
        (\tau;2\zeta_I-P(1)+\tau_0,2\zeta_I-P(1)+\tau_1,-2\zeta_I+P(2)-\tau_2,\dots,\tau_n) & (i=2),\\[5pt]
        (\tau;\tau_0,\dots,2\zeta_I-P(i-1)+\tau_{i-1},-2\zeta_I+P(i)-\tau_i,\dots,\tau_n) & (3\leq i\leq n-2),\\[5pt]
        (\tau; \tau_0,\dots,2\zeta_I-P(n-2)+\tau_{n-2},-2\zeta_I+P(n)-\tau_n,-2\zeta_I+P(n)-\tau_{n-1})& (i=n-1),\\[5pt]
        (\tau; \tau_0,\dots,2\zeta_I-P(n)+\tau_n,-2\zeta_I+P(n)+\tau_{n-1}) & (i=n) \, .
    \end{cases}
\end{equation}} 

Finally, the permutations act on the couplings as

\begin{equation}\label{eq:permutationsD}
    s_{i} : \vec{\tau} \longmapsto
    \begin{cases}
        (\tau; \tau_0 , \, - \tau_1 , \, \tau_2 + \tau_1  , \, \tau_3 , \, \ldots \, , \tau_n) & (i=1),\\[5pt]
        (\tau;\tau_0 + \tau_2 , \, \tau_1 + \tau_2 , \, - \tau_2 , \, \tau_3 + \tau_2 , \, \tau_4 , \,  \ldots \, , \tau_n) & (i=2),\\[5pt]
        (\tau; \tau_0, \, \ldots \, , \tau_{i-1} + \tau_i , \, - \tau_i , \, \tau_{i+1} + \tau_i , \, \tau_{i+2} , \, \ldots  \tau_n) & (3\leq i\leq n-2),\\[5pt]
        (\tau; \tau_0, \, \ldots \, , \tau_{n-4} , \, \tau_{n-3} + \tau_{n-2} , \, - \tau_{n-2} , \,  \tau_{n-1} + \tau_{n-2} , \, \tau_{n} + \tau_{n-2} )& (i=n-2),\\[5pt]
        (\tau; \tau_0, \, \ldots \, , \tau_{n-3} , \, \tau_{n-2} + \tau_{n-1} , \, - \tau_{n-1} , \, \tau_{n} ) & (i=n-1) \, .
    \end{cases}
\end{equation}

The choice in \cref{eq:couplDn} makes manifest the relation between the couplings and the root lattice of the affine $\widehat{D}_n$ algebra and indeed the above transformations generate the automorphisms of the affine root system. 
This can be checked explicitly by writing the deck transformations associated with $\pi_1(\mathbb{T}^2)$ in terms of the $R_I$ generators. The set of $t^{(\tau)}_i$, $s_i$, $R_A$ and $R_D$ can be matched with the generators of the automorphism group of the affine $\widehat{D}_n$ algebra, comprising the Weyl group, see for example \cite{DiFrancesco:1997nk}.

\subsection{Global variants and dualities}\label{sec:globalvariants}

We now address higher form symmetries and duality symmetries that can arise in these quiver theories, discussing in particular how the mapping class group of the Riemann surface used to construct these theories plays a crucial role in both of these aspects.

Let us start by discussing the $\widehat{A}_{n-1}$ case. Recall that this theory can be obtained via a class $\mathcal{S}$ construction, wrapping M5-branes on a torus with $n$ punctures. If all the punctures are regular, one has a 1-form symmetry\footnote{More in general, the worldvolume theory of $k$ M5-branes wrapping a Riemann surfaces $\Sigma_{g,n}$ of genus $g$, with $n$ regular punctures, has a $\mathbb{Z}_k^g$ 1-form symmetry.} \cite{Bah:2018jrv, Bah:2019jts, Bhardwaj:2021pfz, Bhardwaj:2021mzl, Garding:2023unh}. This picture provides a  clear understanding of how the duality group acts on the 1-form symmetry, as follows.

Indeed, the generators of the 1-form symmetry correspond to non-trivial homology 1-cycles of the Riemann surface \cite{Tachikawa:2013hya}. For example, if the Riemann surface is an elliptic curve with underlying smooth surface the torus, the two generators of $H_1(T^2,\mathbb{Z})$ correspond to the symmetry operators capturing the 1-form “electric" and “magnetic" symmetries, which are mutually dual.\footnote{The homology of the Riemann surface encodes all possible global variants of a given theory. In order to specify a particular one, one needs to choose a Lagrangian sublattice of the whole charge lattice of the theory \cite{Bashmakov:2022uek}.} Correspondingly, the usual $A$ and $B$ cycles of the torus form a symplectic basis of $H_1(T^2,\mathbb{Z})$ with respect to the intersection pairing. Because of this, the mapping class group of the surface can act non-trivially on global variants: any transformation which shuffles the generators of $H_1(T^2,\mathbb{Z})$, will, in general, also change the global variant of the theory one considers.
    
The simplest example of this is $\mathcal{N}=4$ $\mathfrak{su}(k)$ SYM, obtained by compactifying the 6d $\mathcal N=(2,0)$ SCFT of type $A_{k-1}$ on an elliptic curve without punctures. Every global variant of the theory has a non-trivial 1-form symmetry \cite{Aharony:2013hda} (see \cite{Bashmakov:2022uek, Antinucci:2022cdi} for the class $\cal S$ perspective). For example, when the gauge group is $\mathrm{SU}(k)$ the 1-form symmetry is purely electric $\mathbb{Z}_k^{(1)}$, and the charged objects are the Wilson loops. On top of that, $\mathcal{N}=4$ SYM enjoys Montonen--Olive duality, exchanging “electric" and “magnetic" degrees of freedom, most notably Wilson and 't Hooft loops. These two facts are captured in the class $\mathcal{S}$ realization of the theory: the 1-form symmetry descends from the reduction of the 2-form symmetry of the 6d SCFT on one of the cycles of the torus, and Montonen--Olive duality is embodied as modular transformations of the elliptic curve, i.e.~the MCG of the underlying torus. In particular, an $S$-duality transformation exchanges the two cycles of the torus, and maps for example $\mathcal{N}=4$ $\mathrm{SU}(k)$ SYM at gauge coupling $\tau$ to $\mathcal{N}=4$ $\mathrm{PSU}(k)_0$ SYM at gauge coupling $-1/\tau$, with the notation of \cite{Aharony:2013hda}. This is one version of Langlands duality \cite{Kapustin:2006pk}.
    
It turns out that the presence of regular punctures on the Riemann surface of the class $\mathcal{S}$ construction plays no role as far as the 1-form symmetry is concerned. This follows from the fact that the additional 1-cycle introduced by each puncture has trivial intersection pairing with any other 1-cycle \cite{Bah:2019jts, Bhardwaj:2021mzl}. This means that only the $\mathrm{SL}(2,\mathbb{Z})$ subgroup of the mapping class group of $\Sigma_{1,n}$, more precisely, its $S$ transformation, can shuffle global variants of the theory, while the other mapping classes have no effect on the 1-form symmetry of the theory, despite inducing genuine dualities.

The $\widehat{D}_n$ case deserves more attention. Indeed, the Riemann surface in this case is an orbifold, for which the notions of intersection pairing and integer 1-cycles are subtle. Even before establishing what is the 1-form symmetry, we can ask what are the dualities that can change the global structure of these theories. By analogy with the $\widehat{A}_{n-1}$ case, we assume that the change in global structure takes place only for the transformations for which $\tau \to -1/\tau$. The only element of the mapping class group that acts on $\tau$ in this way is $S$. We thus conclude that also in the $\widehat{D}_n$ case only the $\mathrm{SL}(2,\mathbb{Z})$ subgroup of the mapping class group can change global variants.

In order to establish what is the 1-form symmetry of these theories, let us again start from the $\widehat{A}_{n-1}$ case. From field theory, it is clear that in the $\mathrm{SU}(k)^n$ global variant, the $\mathbb{Z}_k$ 1-form symmetry acts on the $k$ non-trivial Wilson lines, which are obtained by identifying the Wilson lines of each $\mathrm{SU}(k)$ group due to the presence of dynamical bifundamental matter fields. Performing the $S$ operation, the Wilson lines become $k$ non-trivial 't Hooft lines, meaning that the new global variant is $\mathrm{SU}(k)^n/\mathbb{Z}_k$, which is alternatively obtained by gauging the 1-form symmetry. For the $\widehat{D}_n$ quiver with $\mathrm{SU}$ gauge groups, the bifundamentals are such that again there are only $k$ non-trivial independent Wilson lines, so that the 1-form symmetry is still $\mathbb{Z}_k$.\footnote{More generally, any balanced $\mathcal N=2$ quiver theory without flavor nodes has a $\mathbb{Z}_k$ 1-form symmetry, where $\mathrm{SU}(k)$ is the smallest gauge group in the quiver.} As argued above, the action of $S$ on the global structure of the gauge group is then exactly as in the $\widehat{A}_{n-1}$ case, i.e.~the same as gauging the $\mathbb{Z}_k^{(1)}$.

\subsubsection{Duality symmetries and orbits of marked points}\label{sec:Andualitysymm}

We now describe how dualities of the field theory for $\widehat{A}_{n-1}$ and $\widehat{D}_{n}$ quiver theories can enhance to symmetries at specific points of the conformal manifold. This is because, as we have just seen, some dualities change the global structure of the gauge group, but the latter change can be undone by gauging the 1-form symmetry. Hence if the duality leaves the coupling invariant, the combination of duality and gauging becomes a symmetry of a specific theory. Since a gauging is involved, these duality symmetries are in general non-invertible \cite{Kaidi:2021xfk,Choi:2021kmx,Choi:2022zal}. 

For specific values of the $\tau_i$, there might exist a non-trivial subgroup of the MCG leaving the field theory unchanged. The simplest example is the $S$ transformation of $\mathcal{N}=4$ $\mathfrak{su}(k)$ at $\tau=i$, which leaves the local dynamics invariant, but changes the global variant. One can then recover the original theory by gauging the 1-form symmetry.

In the quiver theories of our interest, the same kind of symmetry transformation can be constructed. One starts by considering mapping classes which fix the configuration of marked points. When such operations act non-trivially on the global structure of the theory, one can compensate them by gauging (a subgroup of) the 1-form symmetry. As in the $\mathcal{N}=4$ case, these two combined operations will, in general, lead to non-invertible symmetries.\footnote{It can happen that no discrete gauging is needed after the action of the MCG, since the global variant may be preserved. When this is the case, the duality symmetry is invertible. If a duality symmetry is non-invertible in every global variant, it is called intrinsically non-invertible, see \cite{Bashmakov:2022uek} for further details.}

The part of the MCG that acts non-trivially on the global variants of the theory is the modular one, moreover only the finite subgroups of $\mathrm{SL}(2,\mathbb{Z})$ that stabilize the coupling $\tau$ can lead to (non-invertible) symmetry defects. These subgroups are cyclic of order $2$, $3$, $4$ or $6$ and are generated by $S^2, S^3T, S$ and $ST$ respectively.\footnote{Another common choice for an element of order $3$ is $ST^{-1}$; however, $ST^{-1}$ fixes $\exp(\pi i/3)$ rather than $\exp(2\pi i/3)$.} 
The transformation $S^2$ leaves $\tau$ unchanged, and therefore can be a symmetry for any choice of $\tau$, depending on the location of the punctures. Meanwhile, $S$, $S^3T$ and $ST$ can be symmetries only for fixed values of $\tau$: $\tau=i$ for $S$ and $\tau=\exp(2 \pi i / 3)$ for both $ST$ and $S^3T$. These transformations will be actual symmetries of the theory, depending on the position of the punctures, as we will discuss shortly.

Before proceeding, a comment is due concerning the terminology. The term `duality' refers in general to the type of relations between theories that are the object of the present paper. However, more specifically, `duality' often refers to the particular action $S$ on the theory space that fixes $\tau=i$. Now, as just emphasized above, such action is actually of order 4, since while $S^2$ sends any $\tau$ to itself, it acts as charge conjugation on the spectrum. As we will see, it permutes the punctures in the cases of our interest. Similarly, `triality' usually refers to both $ST$ and $S^3T$ because they send any $\tau$ to itself after acting three times, but actually their action on the spectrum is respectively of order 6 and 3. Though we will not use `tetrality' instead of duality, since there is no distinction to be made, in the case of triality we will use the term `hexality' when it is important to stress that we are referring to the action which is of order 6 on the spectrum.

\paragraph{Non-invertible defects of $\mathcal N=2$ $\widehat{A}_{n-1}$ quiver SCFTs.}

Let us consider the standard fundamental cell $\mathcal F$ in $\mathbb{C}$ for the torus $E_{\tau=i}$ which is invariant under the action of $S$, i.e. the parallelogram $\{0,1,i,1+i\}$. A point $p\in\mathcal F$ is mapped to $p/i$, which amounts in a clockwise $\pi/2$ rotation with respect to the origin. The point is now out of the fundamental cell $\mathcal F$, as in \cref{fig:a2example}. We can then use the deck transformation $t$ to bring it back to $\mathcal F$:
\begin{equation}\label{eq:tSactionp}
    p \longrightarrow \frac{p}{i}+i = \frac{p-\left(\frac{1+i}{2}\right)}{i}+\left(\frac{1+i}{2}\right)~.
\end{equation}
The way in which the right-hand side is written makes explicit that $t\circ S$ acts as a $(-\pi/2)$-rotation about the center $(1+i)/2$ of $\mathcal F$.

\begin{figure}[h!]
    \centering
    \includegraphics[]{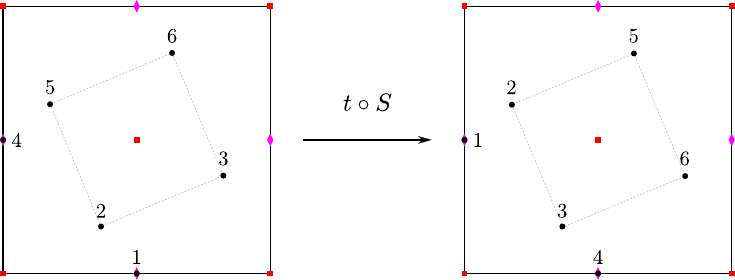}
    \caption{$(t\circ S)$-invariant configuration of points ($\widehat{A}_{n-1}$ case).}\label{fig:SselfdualconfigSection1}
\end{figure}

The points in $\mathcal F$ split in orbits under the action of $t\circ S$, where $t$ here denotes the combination of the deck transformations $t$ for all the marked points. Generic orbits consist of four points; an example is the orbit $\{2,3,5,6\}$ shown in \cref{fig:SselfdualconfigSection1}, in which the points are permuted by $t\circ S$ as $2\rightarrow 5 \rightarrow 6 \rightarrow 3 \rightarrow 2$, which in standard cycle notation reads $(2~5~6~3)$. Apart from the generic orbits, there is one orbit of size two depicted by the purple rhombi in \cref{fig:SselfdualconfigSection1}--the points $1$ and $4$ form an orbit of size two denoted $(1~4)$--and two orbits of size one, depicted as red squares. This way to represent points with non-trivial stabilizers is standard in the theory of wallpaper groups; in the present case, configurations of marked points in $E_i$ invariant under $t\circ S$ have as group of symmetries the wallpaper group denoted $p4$ (in crystallographic notation).

With the labeling on the left of \cref{fig:SselfdualconfigSection1}, the position of the marked points 3, 5 and 6 is determined by the one of 2 and the requirement that the configuration is invariant under $t\circ S$:
\begin{align}\label{eq:orbitsize4}
    p_3 = i p_2 + 1 \; , \qquad p_5 = -i p_2 + i \; , \qquad p_6 = -p_2 + i + 1 \; .
\end{align}

The combined operation $t \circ S$ permutes the punctures accordingly to their orbits under the $(-\pi/2)$-rotation. One can recover the original configuration by a suitable permutation $\sigma$; for example in \Cref{fig:SselfdualconfigSection1} one has $\sigma = (2~3~6~5)(1~4)$, or in terms of $s_i$ it reads $\sigma = s_3 s_2 s_1 s_4 s_3 s_2 s_3 s_5$. Therefore, the combination $\mathcal{D} = \sigma \circ t^{(i)} \circ S$ maps the original theory to itself up to a discrete gauging of the 1-form symmetry acting on the global variants of the theory. In this way one constructs non-invertible duality defects of $\mathcal N=2$ $\widehat{A}_{n-1}$ quivers SCFTs akin to those of \cite{Damia:2023ses}, for each configuration of punctures invariant under $t\circ S$. Such configurations of punctures necessarily split in orbits of size four, two and one. Discrete gauging of the one-form symmetry ensures that this duality enhances to a non-invertible symmetry of the field theory. 

\begin{figure}[h!]
    \centering
    \includegraphics[]{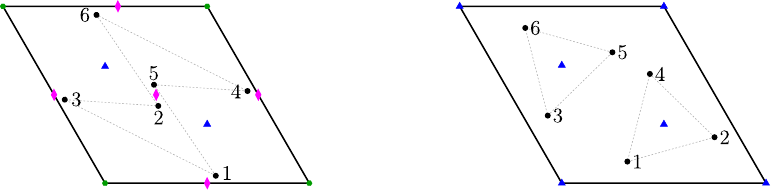}
    \caption{Left (resp. right): $ST$-invariant (resp. $S^3T$-invariant) configuration of points ($\widehat{A}_{n-1}$ case).}\label{fig:U-Uselfdualconfig}
\end{figure}

One can repeat the reasoning replacing $S$ by 
\begin{equation}\label{eq:ST}
    ST = \left(\begin{array}{cc}
    0 & -1 \\ 1 & 1 \end{array}\right)~~~\text{or}~~~S^3T = \left(\begin{array}{cc}
    0 & 1 \\ -1 & -1 \end{array}\right)~,
\end{equation} 
which are respectively of order 6 and 3 in $\mathrm{SL}(2,\mathbb{Z})$. Both $ST$ and $S^3T$ fix $\tau=\exp(2i\pi/3)$, and they act on the marked points as
\begin{equation}
    ST : p\longmapsto p\exp(-i\pi/3)~~~\text{and}~~~ S^3T : p\longmapsto p\exp(2i\pi/3)~,
\end{equation}
that is, as a rotations by $-\pi/3$ or $2\pi/3$, respectively. As before, one can compose $ST$ and $S^3T$ with appropriate deck transformations, to ensure that points in the standard fundamental cell of $E_\tau$--the parallelogram $\{0,1,\tau,\tau+1\}$--are mapped to points of the same fundamental cell. One finds that the generic orbits for $ST$ are of order 6, and that there is one non-generic orbit of size 3, one of size 2 and one of size 1. This is depicted on the left of \cref{fig:U-Uselfdualconfig}, where as before purple rhombi depict the points whose stabilizers are of order 2, whereas blue triangles and green hexagons correspond to those whose stabilizers are of order 3 and 6, respectively. The corresponding wallpaper group is $p6$. The generic orbit of size six shown on the left of \cref{fig:U-Uselfdualconfig} is permuted by $ST$ as the cycle $(1~3~2~6~4~5)$. 

Conversely, generic orbits for $S^3T$ are of size three, and there are three non-generic orbits of size one and with stabilizer of order 3, depicted as blue triangles on the right of \cref{fig:U-Uselfdualconfig}. The corresponding wallpaper group is $p3$. The configuration of points shown there splits in two regular orbits: $(1~4~2)(3~6~5)$.

Again combining the action of $ST$ (resp. $S^3T$) composed with a deck transformation, with an appropriate permutation and a discrete gauging of the one-form symmetry, one ends-up with a comprehensive description of hexality (resp.~triality) non-invertible defects for $\mathcal N=2$ $\widehat{A}_{n-1}$ quivers SCFTs. 

\paragraph{Non-invertible defects of $\mathcal N=2$ $\widehat{D}_{n}$ quiver SCFTs.}

In the case of $\mathcal N=2$ $\widehat{D}_n$ quiver SCFTs, we can apply readily the same method to determine non-invertible defects. Let us discuss the specific example of the $\mathcal N=2$ $\widehat{D}_4$ quiver SCFT at $\tau=i$, whose conformal manifold is described by configurations of four punctures in the lower half of the fundamental cell $E_i$, i.e. Im$(p_i) \leq 1/2$.

\begin{figure}[h!]
    \centering
    \includegraphics[]{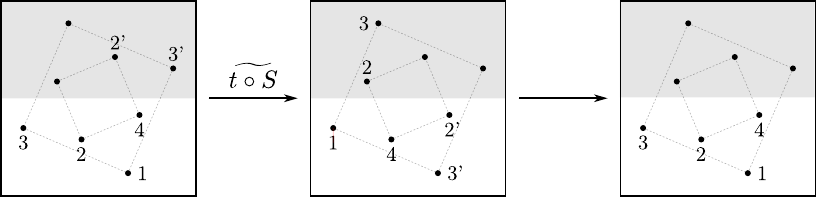}
    \caption{Construction of an non-invertible duality defect in the $\mathcal N=2$ $\widehat{D}_4$ quiver theory}\label{fig:SselfdualconfigD}
\end{figure}

Let $p_1,p_2,p_3$ and $p_4$ denote the position of the marked points $1,2,3$ and $4$ respectively, on the left of \cref{fig:SselfdualconfigD}. We gather them in the tuple $(p_1,p_2,p_3,p_4)$. Recall that the position of the $i'$-th image is determined by the position of the $i$-th marked point:
\begin{equation}
    p_{i'} := q_i^{-1,-1} = -p_i+1+i~. 
\end{equation}

Under $t\circ S$ one has
\begin{equation}
    (p_1,p_2,p_3,p_4) \longrightarrow (-ip_1+i,-ip_2+i,-ip_3+i,-ip_4+i) = (p_3,p_{4'},p_{1'},p_2)~.
\end{equation}

We now apply $R_{D,2}$ and $R_{D,3}$ in order to have all marked points in the desired region of the fundamental cell
\begin{equation}
    R_{D,2} R_{D,3}(p_3,p_{4'},p_{1'},p_2) = (p_3,p_4,p_1,p_2)~.
\end{equation}
Lastly, via a combination of the permutations $s_i$, one can restore the starting configuration of punctures. In the example of \Cref{fig:SselfdualconfigD}, one can take:
\begin{equation}
    s_2s_1s_3s_2: (p_3,p_4,p_1,p_2) \longrightarrow (p_1,p_2,p_3,p_4)~.
\end{equation}
All in all, the non-invertible duality defect is obtained as the operation $s_2s_1s_3s_2 \circ R_{D,2} \circ R_{D,3} \circ t \circ S$ combined with an appropriate discrete gauging of the one-form symmetry.

This procedure generalizes to all $\mathcal N=2$ $\widehat{D}_n$ quiver SCFTs and duality defects, with corresponding wallpaper group $p4$, or triality defects, with corresponding wallpaper group $p6$. Note that by construction, $S^2$ is a symmetry of any $\widehat{D}_n$ configuration of punctures, which implies in particular that triality defects are necessarily of order $6$ and not $3$.

\subsection{Duality group of \texorpdfstring{$\widehat{E}_n$}{En} quiver SCFTs}\label{sec:enduality}

We have seen that the duality group of $\mathcal N=2$ $\widehat{A}_{n-1}$ and $\widehat{D}_n$ quiver SCFTs contains the group of automorphisms of the corresponding affine root system: each coupling constant is naturally associated to a positive simple root of the corresponding affine Lie algebra. The global coupling $\tau$ is defined as\footnote{Here $\tau_i$ are the coupling constants of the single nodes and $n_i$ are the ranks of the nodes in the McKay graph.}
\begin{equation}
        \tau = \sum_i n_i \tau_i 
\end{equation}
and corresponds to the imaginary root $\delta$ of the affine root system.

The full duality group is generated by the automorphism group of the affine root system together with a copy of $\mathrm{SL}(2,\mathbb{Z})$ acting on $\tau$ as:
\begin{align}\label{eq:DualityGroupMassesAn}
\begin{split}
     S &: (\tau,\tau_0,\dots,\tau_n) \mapsto \left(-\frac{1}{\tau},\frac{\tau_0}{\tau},\dots,\frac{\tau_{n-1}}{\tau},\frac{\tau_n}{\tau}-1-\frac{1}{\tau}\right)~,\\
     T &:  (\tau,\tau_0,\dots,\tau_n) \mapsto  (\tau+1,\tau_0,\dots,\tau_{n-1},\tau_n+1)~.
\end{split}
\end{align}
Such modular transformations of the parameter $\tau$ were argued to exist from the underlying class $\cal S$ construction.

It is natural to conjecture that this analysis extends to the $\widehat{E}_{6,7,8}$ quiver gauge theories. These can be constructed as worldvolume theories of D3-branes transverse to $\mathbb{C}^2/\Gamma_{E_n} \times \mathbb{C}$ singularities, where $\Gamma_{E_n}$ is the corresponding finite subgroup of $\mathrm{SU}(2)$.
More precisely, we conjecture that the duality group of these theories is the semi-direct product of $\mathrm{SL}(2,\mathbb{Z})$ with the affine Weyl group, further centrally coextended by the automorphisms of the affine Lie algebra, acting on $\tau$ as in \cref{eq:DualityGroupMassesAn}. From this one can in principle derive the action of the duality group on the global variants of the theory, and hence construct non-invertible defects for suitable configurations of couplings.

Unlike in previous cases, there is no known class 
$\mathcal S$ realization of these theories—at least to our knowledge—so we lack direct methods to test our arguments. This analysis might actually pave the way for discovering explicit class $\mathcal S$ realizations, or generalizations thereof, of $\widehat{E}_n$ quiver theories.

\section{Duality group of the mass deformed theory}\label{sec:MassDefN1}

In the previous section, we have constructed the duality group of $\mathcal{N}=2$ $\widehat{A}_{n-1}$ and $\widehat{D}_n$ quiver gauge theories from class $\mathcal{S}$ arguments. We devote this section to the above quiver theories mass deformed to $\mathcal{N}=1$, exploiting the class $\mathcal{S}$ setup while close in spirit  to \cite{Argyres:1999xu}.

\subsection{Duality group of mass deformed \texorpdfstring{$\widehat{A}_{n-1}$}{A(n-1)}} \label{sec:MassDefAn}

In the following, we revisit the duality group of $\mathcal{N}=1$ theories obtained as mass deformations of $\mathcal{N}=2$ $\widehat{A}_{n-1}$ quiver gauge theories with gauge group $\mathrm{SU}(k)^n$, along the lines of \cite{Halmagyi:2004ju}. We will consider mass deformations of the form
\begin{align}\label{eq:mass_supo}
    \Delta W = \sum_{i=1}^{n} \frac{m_i}{2} \phi_i^2 \; ,
\end{align}
which lead to $\mathcal{N}=1$ SCFTs \cite{Leigh:1995ep,Franco:2015jna,Fazzi:2019gvt}, whose duality groups are induced by the original $\mathcal{N}=2$ theory. 

The mass deformed theory is specified by the masses $(m_1, m_2, \ldots, m_{n})$. As in \cref{sec:N2An}, the strategy to construct the duality group consists in uplifting the associated type IIA elliptic model to M-theory, where the theory can be fully described geometrically.

Recall from \cref{tab:braneelliptic} that the starting type IIA setup consists of $k$ D4-branes on a circle of radius $R_6$ intersecting $n$ NS5-branes along the transverse direction. The low energy theory on the D4-branes is a $4d$ $\mathcal{N}=2$ $\widehat{A}_{n-1}$ quiver gauge theory, where the VEVs of the complex adjoint scalars in $\mathcal N=2$ vector multiplets parameterize the position of the D4-branes along $(x^4,x^5)$. Let: 
\begin{align}
    \begin{split}
    u := x^4 + i x^5 \; , \\
    v := x^7 + i x^8 \; .
    \end{split}
\end{align}
In this setup, the mass deformation we are interested in can be induced by tilting the NS5-branes relatively to each other in the complex $(u,v)$-plane. More precisely, if two adjacent NS5-branes are not parallel then any displacement of the center of mass of a D4-brane stretched between them changes the minimal length of the D4-segment, therefore the D4 are no longer free to move along the NS5's. From the point of view of the field theory, this means that some flat direction has been lifted. 
This lifting can be achieved, in first approximation, by adding  mass terms to the adjoint scalars \cite{Barbon:1997zu}.
One can see that, when the relative angle between two adjacent NS5-branes is small, the mass is directly proportional to the angle \cite{Barbon:1997zu}. In this limit, to which we will refer as the \textit{limit of small masses}, one can identify the small mass with the ones in \cref{eq:mass_supo} \cite{Halmagyi:2004ju}. The $\mathcal{N}=1$ SCFT obtained by integrating them out lives at scales much smaller than the masses in  \cref{eq:mass_supo}.

We now proceed with the uplift to M-theory, where we introduce the elliptic curve $E_\tau$, parameterized by the complex coordinate $w = x^{10} + i x^6$:
\begin{align}
    w \sim w + q + \tau \ell~,~~~q, \ell \in\mathbb{Z}~,
\end{align}
as outlined in \cref{sec:N2An}. The $k$ D4-branes become M5-branes wrapping $E_{\tau}$, whereas the NS5-branes lift to marked points. However, since each NS5-brane corresponds to a specific complex line in the $(u,v)$-plane, we can assign to them the point $[u_i:v_i] \in \mathbb{CP}^1$ which corresponds to the latter.\footnote{For example, the original $\mathcal N=2$ case in which all NS5-branes extend along $u$ is given by $[u_i:v_i]=[1:0]$ $\forall i$.} Thus, the M-theory uplift is encoded in the data of $(p_i,[u_i:v_i])\in E_\tau\times\mathbb{CP}^1$.

In the limit of small masses, the lines of homogeneous coordinates $[u_i:v_i]$ are all close to $[1:0]$, i.e. they are all in the complex chart $\mathbb{C}\simeq\{u\neq 0\}\subset \mathbb{CP}^1$ and one can always rewrite $[u_i:v_i]=[1:z_i]$ with $z_i\in\mathbb{C}$, where now the small mass limit can be formally expressed as $z_i \to 0$ $\forall i$. As a consequence of supersymmetry, since the superpotential needs to be holomorphic in the fields, the mass parameters $m_i$ must be holomorphic in $z_i$ and vanish when all branes are parallel, i.e. when $z_i$ approaches $z_{i+1}$ \cite{Halmagyi:2004ju}. This implies:
\begin{align}\label{eq:massesangles}
    m_i = z_{i+1} - z_i \; , \quad z_i \sim z_{i+n} \; .
\end{align}

Tilting the $n$ NS5-branes in the type IIA setup yields only $n-1$ relative angles, which is at odds with the $n$ mass parameters one can define in field theory. Equivalently, given the definition in \cref{eq:massesangles}, the masses satisfy $\sum m_i = 0$. This apparent paradox can be tackled similarly to what is done in \cite{Witten:1997sc} by considering a non-trivial $\mathbb{CP}^1$-fibration over the elliptic curve $E_\tau$. Indeed, as we will see shortly, if we consider the $z_i$ as sections of a non-trivial fibration over $E_\tau$, we can recover the missing mass deformation in term of a “global mass", $m = \sum m_i$, which vanishes precisely when the fibration trivializes. 

Let us denote $R$ ("Right") and $U$ ("Up") the topologically non-trivial cycles of $E_\tau$ corresponding to $w\rightarrow w+1$ and $w\rightarrow w+\tau$, respectively. Saying that $\mathbb{CP}^1$ is fibered non-trivially over $E_\tau$ means that there can be non-trivial (projective) monodromies along $R$ and $U$, in the form of matrices in $\mathrm{GL}(2,\mathbb{C})$ acting projectively on the fiber coordinate:
\begin{align}
    z \rightarrow \frac{a z + b}{c z + d} \; .
\end{align}

In order for the fibration to preserve $\mathcal N=1$ supersymmetry, the monodromy along either cycles of the torus must preserve the holomorphic 3-form $\Omega = du \wedge dv \wedge dw$, and this implies that the monodromies actually live in $\mathrm{SL}(2,\mathbb{C})$: 
\begin{align}
    M = \left( 
    \begin{matrix}
        a & b \\
        c & d
    \end{matrix}
    \right) \; , \quad a d - b c = 1 \; . 
\end{align}
In field theory the mass deformation can be continuously turned off, which implies that the fibration must be topologically trivial. It is then entirely described by the monodromies $M_R$ and $M_U$ associated to the cycles $R$ and $U$ of $E_\tau$, which provide a representation of the fundamental group of $E_\tau$. Since $\pi_1(E_\tau)$ is abelian, the matrices $M_R$ and $M_U$ must commute, as we would have intuitively expected. 

In the limit of small masses, one can approximate the projective bundle with an affine fibration over $E_{\tau}$. From field theory one expects that the monodromies in $\mathrm{SL}(2,\mathbb{C})$ act as shifts $z_i \to z_i + b$ in this limit. Note that when $z$ is small:
\begin{align}
    \frac{a z + b}{c z + d} \simeq \frac{a d - b c}{d^2} z + \frac{b}{d} + \mathcal{O}(z^2) = \frac{1}{d^2} z + \frac{b}{d} + \mathcal{O}(z^2) \; ,
\end{align}
thus it must be the case that $d=1$. The affine shift is non-trivial when $b \neq 0$, which we now assume. The monodromies $M \in \mathrm{SL}(2,\mathbb{C})$ must therefore be of the form:
\begin{align}
    M = \left( 
    \begin{matrix}
        a & b \\
        \frac{a-1}{b} & 1
    \end{matrix}
    \right) \; . 
\end{align}
One can check that two such generic matrices commute if and only if $a=1$, thus we define:
\begin{align}
            M_R = \left( 
            \begin{matrix}
                1 & b_R \\
                0 & 1
            \end{matrix}
            \right) \; , \quad
            M_U = \left( 
            \begin{matrix}
                1 & b_U \\
                0 & 1
            \end{matrix}
            \right) \; ,
\end{align}
where $b_R$ and $b_U$ are complex numbers characterizing the fibration.\footnote{Our solution is slightly different from the one in \cite{Halmagyi:2004ju}, but it makes more explicit the relation between $M_{R,U}$ and the shifts $z\rightarrow z+\mathrm{constant}$.} The monodromies induce the following transformations:
\begin{align}
    \begin{split}
    &(\tau, b_R, b_U, p_i, z_i) \; \overset{R}{\longrightarrow} \; (\tau, b_R, b_U, p_i + 1, z_i + b_R) \; , \\
    &(\tau, b_R, b_U, p_i, z_i) \; \overset{U}{\longrightarrow} \; (\tau, b_R, b_U, p_i + \tau, z_i + b_U) \; .
    \end{split}\label{eq:gluingfunction}
\end{align}
Allowing the affine fibration over $E_\tau$ to be non-trivial introduces the freedom to change the fiber coordinate $z$ by $(p,z)\rightarrow (p, z + \lambda p)$, such that the shifts of the monodromies in \cref{eq:gluingfunction} are preserved. This ``gauge" symmetry acts\footnote{This action is recovered by asking the action of $M_{R,U}$ on the parameter to match before and after the gauge fixing.} on the parameters of the M-theory setup as: 
\begin{align}
    (\tau, b_R, b_U, p_i, z_i) \overset{f}{\longrightarrow} (\tau, b_R + \lambda, b_U + \tau \lambda, p_i, z_i + \lambda p_i) \; .
\end{align}
One can fix the gauge by imposing $M_R$ to be trivial, i.e. $b_R=-\lambda$ and we can define
\begin{align}
    b_U - b_R \tau = m \; 
\end{align}
to be the global mass.

To conclude, in M-theory the setup is fully specified by the tuple
\begin{align}
    \left( \tau, 0, m, p_i, z_i\right) \; ,
\end{align}
where $m$ and the $z_i$ can be traded for $n$ masses $m_i$:
\begin{align}\label{eq:Anmasses}
    m_i &= z_{i+i} - z_{i} \; , \qquad i = 1, \ldots , n-1 \; , \nonumber \\
    m_n & = m + z_{1} - z_{n} \; , \nonumber \\
    m &= \sum_{i=0}^{n-1} m_i \; .
\end{align}
The above equations show how a non-trivial fibration allows for a tilted configuration with only one mass term, say $m=z_n$. This can also be understood as a consequence of \cref{eq:massesangles}, where we defined $m_n = z_1-z_n$. When we have a non-trivial fibration, the difference between $z_1$ and $z_n$ is computed across the fundamental cell of the torus, thus we should consider not $z_1$, but $M_U(z_1)=z_1+m$ and hence the definition in \cref{eq:Anmasses}. 

We see that the mass deformed theory is now fully specified by the vector $\vec{m}=\left( m; \, m_1, \, \ldots \, , m_{n} \right)$. However, this set of variables depends of the gauge fixing and it is not preserved by $S$, which exchanges the $R$ and $U$ cycles and consequently $b_R$ and $b_U$. Explicitly:
\begin{align}
    (\tau, 0, m, p_i, z_i) \overset{S}{\longrightarrow} \left(-\frac{1}{\tau}, m, 0, \frac{p_i}{\tau}, z_i \right) \; .
\end{align}
Therefore the action of $S$ needs to be followed by another $f$-gauge fixing with $\lambda = - m$, leading to:
\begin{align}\label{eq:Sgaugefixing}
    &(\tau, 0, m, p_i, z_i) \overset{f \circ S}{\longrightarrow} \left(-\frac{1}{\tau}, 0, \frac{m}{\tau}, \frac{p_i}{\tau}, z_i - m \frac{p_i}{\tau} \right) \; .
\end{align}
From now on we assume that $S$ is always post-composed with a suitable $f$-gauge fixing. 

We are now set to describe the duality group of these $\mathcal N=1$ theories. By construction, the action of $T$ and $t_i^{(1)}$ is trivial on the $z_i$, whereas they act on $p_i$ and $\tau$ as in \cref{sec:N2An}. The generator $t_i^{(\tau)}$ moves the punctures along $U$, thus:
\begin{align}
    (\tau, 0, m, p_i, z_i) \overset{t_i^{(\tau)}}{\longrightarrow} (\tau, 0, m, p_i + \tau, z_i + m) \; .
\end{align}
Last, permutations $s_i$ exchange $z_i$ and $z_{i+1}$ as well as $p_i$ and $p_{i+1}$. 

With the masses defined as in \cref{eq:Anmasses}, we can write the action of the generators $\{ S, T, t_i^{(1)}, t_i^{(\tau)} , s_i \}$ of the duality group on the masses as
\begin{align}\label{eq:DualityMasses}
    \begin{split}
            S \; &: \; \vec{m} \; \rightarrow \; \left( \frac{m}{\tau} ; \, m_1 - m \frac{\tau_1}{\tau}, \, \ldots \, , m_i - m \frac{\tau_i}{\tau} , \ldots , m_{n} - m \frac{\tau_{n}}{\tau} + \frac{m}{\tau} \right) \; , \\[5pt]
            t_i^{(\tau)} \; &: \; \vec{m} \; \rightarrow \; ( m; \, m_1 , \, \ldots \, , m_{i-1} + m , m_{i} - m, m_{i+1}, \, \ldots \, , m_{n}) \; , \\[5pt]
            s_i \; &: \; \vec{m} \; \rightarrow \; ( m; \, m_1, \, \ldots \, , m_{i-1} + m_{i} , - m_{i},  m_{i+1} + m_{i} , \ldots , m_{n}) \; , 
    \end{split}
\end{align}
while the remaining generators $T$ and $t_i^{(1)}$ act trivially on $\vec{m}$.

We conclude this section by remarking that the duality group of the mass-deformed theory can be presented similarly to the duality group of the original $\mathcal{N}=2$ theory. The masses behave as roots of the affine $\widehat{A}_{n-1}$ algebra, with the global mass playing the role of the imaginary root. Therefore, the duality group is the extension of the automorphism group of the $\widehat{A}_{n-1}$ algebra by the modular group $\mathrm{SL}(2,\mathbb{Z})$, acting as in \cref{eq:DualityMasses}, while acting as well on the couplings $\tau_i$ as in \cref{eq:mcgtau}.

\subsection{Duality group of mass deformed \texorpdfstring{$\widehat{D}_n$}{Dn}}\label{sec:MassDefDn}

We now address the mass deformations of the $\widehat{D}_n$ theory by a superpotential of the form \cref{eq:mass_supo}, with the masses defining the deformation gathered in a vector $\vec{m}=(m; \, m_0,\dots,m_n)$. As reviewed in \cref{sec:DnUplifts}, this theory admits a type IIA construction in terms of D4s suspended between NS5s, in presence of orientifold ONS5$^-$-planes, \cref{tab:IIAbranes}. In terms of branes, mass deformations are obtained as in \cref{sec:MassDefAn} by tilting the NS5 branes in the $\C^2_{u,v}$ plane. The lift to M-theory then fully unveils the duality group of the deformed theory. 
    
\begin{figure}[h!]
    \centering
    \includegraphics[]{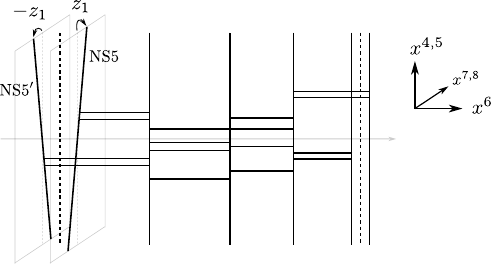}
    \caption{The leftmost NS5 is tilted by $z_1\in\mathbb{C}$ and its image NS5', by $-z_1$.}\label{fig:Dnconfig2}
    \end{figure}

    Each NS5-brane corresponds to a complex line in the $\C^2_{u,v}$ plane, and hence to a point $[u_i : v_i] \in \mathbb{CP}^1$. In the limit of small angles, one can assume that the slope of these lines is close to $[1:0]$, thus one can set $[u_i : v_i]=[1 : z_i]$ where  $z_i \in \mathbb{C}$. The masses can then be expressed in term of the $z_i$, as in \cref{sec:MassDefAn}. The main difference with respect to $\widehat{A}_{n-1}$ theories comes from the orientifold projection: since it maps $x^{7,8}$ to $-x^{7,8}$, tilting an NS5 brane by a complex number $z$ amounts to tilting its image with respect to the $\mathrm{ONS5}^-$-planes by $-z$, at least when the $\C^2_{u,v}$ plane is trivially fibered over the $x^6$-segment; this is depicted in \cref{fig:Dnconfig2} (which builds on the previous \cref{fig:Dnconfig}) for the brane closest to the leftmost $\mathrm{ONS5}^-$-plane, and its image.
    
    Similarly to what was done above in $\widehat{A}_{n-1}$ theories, the masses are expressed in terms of the $z_i$ as:
    \begin{align}\label{eq:dn_mass_zi}
        \begin{split}
        m_0 &= z_2-(-z_1) \, ,\\
        m_{i} &= z_{i+1}-z_{i} \, , \quad i=1, \, \ldots\, , n-1~,\\
        m_n &= +(-z_n)-z_{n-1} \, ,
        \end{split}
    \end{align}
    where we kept the $-$ signs coming from the orientifold projection since they will be relevant in the following. This definition is consistent with the requirement of holomorphy in the $z_i$'s and vanishing of all masses when the branes are parallel.

    The definition in \cref{eq:dn_mass_zi} leads to a vanishing total mass:
    \begin{equation}
        m=m_0+m_1+m_{n-1}+m_n+2 \sum_{i=2}^{n-2} m_i=0~, 
    \end{equation} 
    hence as in $\widehat{A}_{n-1}$ theories it naively seems that there is a missing mass parameter in the brane setup, as compared to the adjoint masses appearing of field theory. Here again, the mismatch is resolved by considering slightly more general brane setups in which $\mathbb{CP}^1_{u,v}$ is allowed to fiber non-trivially over the M-theory pillowcase $E_{\tau}/\mathbb{Z}_2$.
    
    Such a fibration is specified by a representation of the fundamental group $\pi_1(E_\tau/\mathbb{Z}_2)$ into $\mathrm{GL}(2,\mathbb{C})$. Because of the way the orientifolds act on the coordinates $x^{7,8}$, in order to preserve $\mathcal N=1$ supersymmetry the generators $R_A$, $R_{B}$, $R_{C}$ and $R_{D}$ must correspond to matrices of determinant $-1$. Moreover, in the limit of small $z_i$, one expects the monodromies corresponding to $R_l$, $l=A,B,C,D$, to act as $z_i\rightarrow -z_i + b_l$, where the $b_l$ are $i$-independent complex numbers. Recalling that the $R_l$ are involutions, the form of such elementary monodromies is constrained to be: 
        \begin{align}
            M_l = \left(
                \begin{matrix}
                    a_l & b_l \\
                    \frac{1-a_l^2}{b_l} & -a_l
                \end{matrix}
                \right) \; , \quad b_l \neq 0 \; , \quad l = A, B, C, D~,
        \end{align}
    and the other relations eventually yield:
    \begin{align}\label{eq:dngenrel}
        &M_l = \left(
                \begin{matrix}
                    -1 & b_l \\
                    0 & 1
                \end{matrix}
                \right) \; , \quad b_l \neq 0 \; , \quad l = A, B, C, D \nonumber \\[5pt]
        & b_{B} - b_{A} = b_{D} - b_{C} \; .        
    \end{align}
    
    As in \cref{sec:MassDefAn}, allowing non-trivial fibrations over $E_\tau/\mathbb{Z}_2$ introduces an additional freedom in the choice of fiber coordinate. One can do the redefinition $z\rightarrow z+f(p)$ where $p\in E_\tau/\mathbb{Z}_2$ and with $f$ a holomorphic function, however only affine functions $f(p)=\lambda p +\kappa$ preserve the form of the monodromies. Such a coordinate change induces the following transformation:
    \begin{align}
    (\tau, b_A,b_B,b_C,b_D, p_i, z_i) \overset{f}{\longrightarrow} (\tau, b_A , b_B + \lambda ,b_C + \tau \lambda ,b_D + (1+\tau)\lambda, p_i, z_i + \lambda p_i) \; .
    \end{align}
    
    With the physical interpretation in mind, we can fix $\lambda$ in such a way that:
        \begin{align}\label{eq:gaugefixingDn}
            & b_{B} - b_{A} + \lambda = b_{D} - b_{C} + \lambda = 0 \; ,
        \end{align}
    This makes clear that up to redefinition of the fiber coordinate, the fibration depends only on the two parameters $b_A$ and $b_E = b_C + (b_A - b_B) \tau $. With the notation of above, the fibration is defined by the data:
    \begin{align}
        (\tau, b_A , b_E , p_i , z_i )\; ,
    \end{align}
    on which the monodromies act as follows:
    \begin{align}\label{eq:dnmonodromies}
            &(\tau, b_A, b_E, p_i , z_i ) \overset{R_A}{\longrightarrow} (\tau, b_A, b_E, -p_i, - z_i + b_A ) \; , \nonumber \\
            &(\tau, b_A, b_E, p_i , z_i ) \overset{R_B}{\longrightarrow} (\tau, b_A, b_E, - p_i + 1 , - z_i + b_A ) \; , \nonumber \\
            &(\tau, b_A, b_E, p_i , z_i ) \overset{R_C}{\longrightarrow} (\tau, b_A, b_E, - p_i + \tau , - z_i + b_{E} ) \; , \nonumber \\
            &(\tau, b_A, b_E, p_i , z_i ) \overset{R_D}{\longrightarrow} (\tau, b_A, b_E, - p_i + \tau + 1 , - z_i + b_{E} ) \;~.
    \end{align}
    
    In \cref{eq:dn_mass_zi}, we stressed that the tilting of a brane and its image with respect to an orientifold plane are not independent. In M-theory, this amounts to saying that given the tilting $z_i$ of a brane, the tilting of its image with respect to a fixed point of the involution $\mathbb{Z}_2$ is encoded in the image of $z_i$ by the monodromy $M_l$ around that fixed point. In general, the fibration is not trivial, and:
    \begin{align}
        M_l(z_i) = - z_i + b_l \; .
    \end{align}
    This leads to the following generalization of \cref{eq:dn_mass_zi}:
    \begin{align}
        \begin{split}
        &m_0 = z_2 - M_{A/B}(z_1) = z_2 + z_1 - b_A \, ,\\
        &m_{i} = z_{i+1}-z_{i} \, , \qquad i=1,\, \ldots \, , n-1\\
        &m_n = M_{C/D}(z_n)-z_{n-1} = - z_n -z_{n-1} + b_E \, .
        \end{split}
    \end{align}
    \Cref{eq:dn_mass_zi} corresponds to a trivial fibration, for which $b_A=b_E=0$.
    
    The corresponding global mass reads
    \begin{equation}
        m = \sum_i n_i m_i = b_E - b_A
    \end{equation}
    with the $n_i$ the Dynkin labels of affine $\widehat{D}_n$, as in \cref{eq:Dynkinlabels2}. Note that as expected, the global mass $m$ vanishes when the fibration is trivial. 

    We have thus shown that the theories obtained by deforming $\mathcal N=2$ $\widehat{D}_n$ quiver SCFTs by $\mathcal N=1$ preserving masses are fully determined by the set of gauge couplings $\tau_i$ satisfying $\tau=\sum_i n_i \tau_i$, and the set of adjoint masses $m_i$ with $m=\sum_i n_i m_i$. Though the set of couplings and the set of masses play very similar roles in the geometric description of the theories we are interested in, there is an important difference in the way the mapping class group acts on them. Its action on the couplings is given in \cref{sec:DnUplifts}, whereas the one on the masses is described as follows. 
    
    First of all one can note that the action of $T$ is trivial, whereas $S$,\footnote{Composed with a redefinition of the fiber coordinate for the same reason as in \cref{eq:Sgaugefixing}.} acts as
    \begin{align}
        \vec{m} \overset{S}{\longrightarrow} \left( \frac{m}{\tau} ; \, m_0 - m \frac{\tau_0}{\tau}, \, \ldots \, , m_i - m \frac{\tau_i}{\tau} , \, \ldots \, ,  m_n - m \frac{\tau_n}{\tau} + \frac{m}{\tau}\right) \, .
    \end{align}

    Deck transformations act on the masses as in \cref{eq:deckD} with the $m_i$ in place of the $\tau_i$, and with the shifts of the $z_i$ due to the non-trivial fibration taken into account. For example, $R_{I,i}$ maps 
    $\vec{m}$ to:

    {\small
    \begin{equation}\label{eq:deckDmasses}
    \begin{cases}
        (m;m_1+\delta_{C|D,I}m,m_0+\delta_{C|D,I}m,m_2,\dots,m_n) & (i=1),\\[5pt]
        (m;-m_1+\delta_{C|D,I}m,-m_0+\delta_{C|D,I}m,P(2)-m_2-\delta_{C|D,I}m,\dots,m_n) & (i=2),\\[5pt]
        (m;m_0,\dots,-P(i-2)-m_{i-1}+\delta_{C|D,I}m,P(i-1)+m_i-\delta_{C|D,I}m,\dots,m_n) & (3\leq i\leq n-2),\\[5pt]
        (m; m_0,\dots,-P(n-2)+m_{n-2}+\delta_{C|D,I}m,\delta_{A|B,I}m-m_n,\delta_{A|B,I}m-m_{n-1})& (i=n-1),\\[5pt]
        (m; m_0,\dots,m_n+\delta_{A|B,I}m,m_{n-1}+\delta_{A|B,I}m) & (i=n) \, ,
    \end{cases}
\end{equation}}
where $I=A,B,C,D$, where $C|D$ in $\delta_{C|D,I}$ means either $C$ or $D$, and with: 
\begin{equation}
    P(i)=\sum_{k=0}^{i}n_k m_k~.
\end{equation}

Finally the transposition $s_i$ maps the mass vector $\vec{m}$ to:
\begin{equation}\label{eq:permutmassD}
    \begin{cases}
        (m; m_0 , \, - m_1 , \, m_2 + m_1  , \, m_3 , \, \ldots \, , m_n) & (i=1),\\[5pt]
        (m;m_0 + m_2 , \, m_1 + m_2 , \, - m_2 , \, m_3 + m_2 , \, m_4 , \,  \ldots \, , m_n) & (i=2),\\[5pt]
        (m; m_0, \, \ldots \, , m_{i-1} + m_i , \, - m_i , \, m_{i+1} + m_i , \, m_{i+2} , \, \ldots  m_n) & (3\leq i\leq n-3),\\[5pt]
        (m; m_0, \, \ldots \, , m_{n-4} , \, m_{n-3} + m_{n-2} , \, - m_{n-2} , \,  m_{n-1} + m_{n-2} , \, m_{n} + m_{n-2} )& (i=n-2),\\[5pt]
        (m; m_0, \, \ldots \, , m_{n-3} , \, m_{n-2} + m_{n-1} , \, - m_{n-1} , \, m_{n} ) & (i=n-1).
    \end{cases}
\end{equation}
This concludes our analysis of the duality group's action on the masses that define the deformation of $\widehat{D}_n$ quivers.

\section{Moduli space of the mass deformed theory}\label{sec:modulispace}

We have seen in the previous section how the respective MCGs of the $\widehat{A}_{n-1}$ and $\widehat{D}_n$ ${\cal N}=2$ SCFTs act on the mass parameters that one can turn on. Such relevant mass deformations break supersymmetry to ${\cal N}=1$ and trigger an RG flow. In the present section, we ask in all generality what is the moduli space of the theory that is the result of this RG flow. As we will see, such moduli spaces describe geometries which are often a non-trivial fibrations of the geometry described by the ${\cal N}=2$ moduli space.

\subsection{Moduli space of mass deformed \texorpdfstring{$\widehat{A}_{n-1}$}{A(n-1)}} \label{sec:modspaceAn}

We are interested in $\widehat{A}_{n-1}$ quiver theories deformed by $\mathcal{N}=1$ preserving masses, that is:
\begin{align}
    \Delta W = \sum_{i=1}^n \frac{m_i}{2} \phi_i^2 \, ,
\end{align}
where the $\phi_i$ denote the adjoint scalars in the $\mathcal N=2$ vector superfields. 

On general grounds, one expects that the deformed theories flow to interacting $\mathcal{N}=1$ SCFTs \cite{Leigh:1995ep,Fazzi:2019gvt}. One of the simplest examples is the conifold field theory, which is the mass deformation of the $\mathcal N=2$ $\widehat{A}_1$ quiver gauge theory with mass parameters $(m_1,m_2)=(m,-m)$ \cite{Klebanov:1998hh}. Other SCFTs of interest can be obtained from other choices of masses; for example, the Pilch--Warner (PW) point \cite{Pilch:2000ej, Pilch:2000fu, Halmagyi:2004jy, Corrado:2004bz, Benvenuti:2005wi} is also obtained from the $\mathcal N=2$ $\widehat{A}_1$ quiver gauge theory, though with the choice of deformation parameters $(m_1,m_2)=(m,m)$. The moduli space of the former is given by the locus $xy=zw$ in $\mathbb{C}^4$, while the latter's one is the two-fold $\mathbb{C}^2/\mathbb{Z}_2$. The general description of the moduli space of $\mathcal N=1$ deformations of $\mathcal N=2$ quiver gauge theories that we are going to present, will in some cases allow us to argue directly that these theories flow to interacting $\mathcal N=1$ SCFTs.

We first consider general mass deformations of the $\widehat{A}_{n-1}$ quiver gauge theory, with gauge group $\mathrm{U}(1)^n$.\footnote{In the general case where the gauge group is $\mathrm{SU}(k)^n$ for some $k$, the moduli space is generically the $k$-th symmetric product of the abelian one, hence the customary simplification when discussing the moduli space.} The deformed superpotential reads:
\begin{align}
    W_{\mathcal{N}=1} = \sum_{i=1}^{n} \phi_i \left( X_{i,i+1}X_{i+1,i} - X_{i,i-1}X_{i-1,i} \right) + \sum_{i=1}^{n} \frac{m_i}{2} {\phi_i}^2 \; , 
\end{align}
where $i$ is understood modulo $n$. Let
\begin{align}
    &x = \prod_{i=1}^{n} X_{i,i+1} \; , \quad y =  \prod_{i=1}^{n} X_{i,i-1} \; , \nonumber \\ 
    &w_i =  \, X_{i,i-1}X_{i-1,i} \quad \forall \, i \; , \nonumber \\[4pt] 
    &u_i =  \, \phi_i \quad \forall \, i \; ,
\end{align}
be the elementary gauge invariant operators. They are constrained by F-terms equations, which read
\begin{align}
    &X_{i,i+1}X_{i+1,i} - X_{i,i-1}X_{i-1,i} + m_i \phi_i = 0 \; , \nonumber \\
    &X_{i+1,i} \, \phi_{i} - \phi_{i+1} \, X_{i+1,i} = 0 \; , 
\end{align}
for all $i=1,\dots,n$. These lead to the relations: 
\begin{align}
\begin{cases}
    x y = \displaystyle\prod_{k=1}^{n} w_k \; , \\
    u_i = u \; , \\ 
    w_{i+1} - w_{i} = - m_{i} u \; ,
\end{cases}
\end{align}
again for all $i=1,\dots,n$.
The $w_i$'s can be written recursively as
\begin{align}
    w_i = w_1 - \left( \sum_{k=1}^{i-1} m_{k} \right) u = w_1 - t_i u  \; ,
\end{align}
and, since by definition $w_{n+1} = w_1$, we have the constraint
\begin{align}\label{eq:globalmassTimesU}
    \left( \sum_{i=1}^{n} m_i \right) u = m \, u = 0 \; ,
\end{align}
where $m$ denotes the “global mass".

Therefore, denoting $w_1 = w$, the moduli space of the deformed theory is defined by the equations
\begin{align}\label{eq:andeformedmoduli}
    \begin{cases}
        x y = \displaystyle\prod_{k=1}^{n} \left( w - t_k u\right) \; \\
        m u = 0 \; .
    \end{cases}
\end{align}
Note that the second equation imposes either $u=0$ or $m=0$. 

If $m\neq 0$, then $u=0$ and the moduli space is defined by
\begin{align}
    &w_i = w \quad \forall \, i \; , \quad x y = w^n \; , 
\end{align}
i.e. it is the 2-fold $\mathbb{C}^2/\mathbb{Z}_n$. This generalizes the case of the PW fixed point. If rather $m=0$, the moduli space is a 3-fold determined by the partial sums $t_k$. This is analogous to the case for the conifold theory.

From this analysis, we see that the moduli space of the deformed theory is either a two- or a three-fold singularity. In particular, the former is a Du Val singularity of type $A_{n-1}$, while the latter is a compound Du Val, again of type $A_{n-1}$, i.e. $xy=w^n + u g(w,u)$,\footnote{In general, a compound Du Val three-fold is given by the equation $f_{\text{Du Val}}(x,y,w) + u g(w,u)=0$ in $\mathbb{C}^4$.} for some polynomial $g(w,u)$. The two-fold case is less explored in the literature, only for specific examples there are results showing that the mass deformation we are interested in leads to an interacting SCFTs\footnote{As a consistency check, $a$-maximization can be performed to test for a possible violation of the unitarity bound. In the theories under consideration, the deformation introduces quartic interactions analogous to those in the conifold and Pilch–Warner cases, and the resulting $R$-charges are found to satisfy the unitarity bound. This result also holds for the deformed $\hat{D}_n$ models discussed in the following subsection.} \cite{Khavaev:1998fb, Corrado:2002wx, Lunin:2005jy, Butti:2006nk}. On the other hand, in the three-fold case, it has been proven that the deformations under consideration always lead to interacting SCFTs \cite{Fazzi:2019gvt}.

Finally, let us give a different perspective on the IR moduli space in \cref{eq:andeformedmoduli}. In \cite{Lindstrom:1999pz}, a graphical tool called ``bug calculus'' is exploited in order to deform the algebraic curves of ADE singularities with FI terms $b_i$, associated to each node of the extended Dynkin diagram. The singularity is deformed by a versal deformation that depends on the FI parameters $b_i$. In the case of $\widehat{A}_{n-1}$ quiver, one finds
\begin{align}\label{eq:AnFImodulispace}
    x y = \prod_{k=1}^{n} \left[ w - \left( \sum_{j=1}^{k+1} b_j \right) \right] \; ,
\end{align}
with the condition $b_1+\dots+b_n = 0$, which closely resembles \cref{eq:andeformedmoduli}. 

The F-terms equations, after mass deformation, have the same form of the gauge invariants constructed in \cite{Lindstrom:1999pz}, provided the correspondence $b_i \leftrightarrow m_i \phi_i$. A formal correspondence can be established if we deform the $\mathcal{N}=2$ superpotential with complex FI terms,
\begin{align}\label{eq:AndeformedPolonyi}
    W_{\mathcal{N}=1} = \sum_{i=1}^{n} \phi_i \left( X_{i,i+1}X_{i+1,i} - X_{i,i-1}X_{i-1,i} \right) + \sum_{i=1}^{n} b_i \phi_i \; .
\end{align}
After applying the “bug calculus" procedure, we can now trade the $b_i$ for  $m_i \phi_i$ to get to
\begin{align}
    b_i &\mapsto m_i \phi_i \; , \nonumber \\
    \sum_{i=1}^{n} b_i = 0 &\mapsto \sum_{i=1}^{n} m_i \phi_i = 0 \; .
\end{align}
As discussed above, the F-terms requires $\phi_i = u$, thereby obtaining the condition $m\,u=0$. 

This approach will be used in the next section to get the moduli space of deformed $\widehat{D}_n$ quiver theories.
    
\subsection{Moduli space of mass deformed \texorpdfstring{$\widehat{D}_n$}{Dn}}\label{sec:modspaceDn}

The analysis of the previous section can be repeated for the $\widehat{D}_n$ theory.\footnote{The same can also be done for $\widehat{E}_n$-quivers, the resulting moduli space is either $\mathbb{C}^2/\Gamma_{E_n}$, for non-vanishing global mass, or a compound Du Val of type $E$.} Let us start by considering the superpotential for the $\widehat{D}_n$ theory with the addition of masses for the adjoint fields 
\begin{align}
    W_{\mathcal{N}=1} &= \sum_{i=0,1} \phi_i X_{i,2} X_{2,i} + \sum_{j=n-1,n} \phi_{j} X_{j,n-2} X_{n-2,j} + \phi_2 \left( X_{2,0} X_{0,2} + X_{2,1} X_{1,2} + X_{23} X_{32} \right) \nonumber \\
    & + \phi_{n-2} \left( X_{n-2,n-1} X_{n-1,n-2} + X_{n-2,n} X_{n,n-2} - X_{n-2,n-3}X_{n-3,n-2} \right) \, \nonumber \\
    & + \sum_{l=3}^{n-3} \phi_l \left( X_{l,l+1}X_{l+1,l} - X_{l,l-1}X_{l-1,l} \right) +\sum_{i=0}^{n} \frac{m_i}{2} \phi_i^2 \; ,
\end{align}
where we refer to \cref{fig:Dnconfig} for the index conventions of the fields.

As before, we start by considering a theory with abelian gauge factors for the external nodes and $\mathrm{U}(2)$ for the internal ones. Analogously to the previous section, the moduli space of the non-abelian theory can be recovered from the symmetry product of $k$ copies the this moduli space.
The F-terms for the chiral fields are
\begin{align}\label{eq:ftermX}
\begin{cases}
    & \phi_2 X_{2,i} = - X_{2,i} \phi_i \; , \quad i = 0, \, 1  \; ,  \\
    & X_{i,2} \phi_2 = - \phi_i X_{i,2} \; , \quad i = 0, \, 1  \; ,  \\
    & \phi_{n-2} X_{n-2,j} = - X_{n-2,j} \phi_{j} \; , \quad j = n-1, \, n \; ,  \\
    & X_{j,n-2} \phi_{n-2} = - \phi_{j} X_{j,n-2} \; , \quad j = n-1, \, n  \; ,  \\
    & \phi_l X_{l,l+1} = X_{l,l+1} \phi_{l+1} \; , \quad l = 2, \, \ldots \, , n-2 \; , \\
    & X_{l+1,l} \phi_{l} = \phi_{l+1} X_{l+1,l} \; , \quad l = 2, \, \ldots \, , n-2 \; ,
\end{cases}
\end{align}
while for the adjoint fields we have
\begin{align}\label{eq:ftermphi}
\begin{cases}
    & X_{i,2} X_{2,i} = - m_i \phi_i \; , \quad i = 0, \, 1  \; ,  \\
    & X_{j,n-2} X_{n-2,j} = - m_j \phi_j \; , \quad j = n-1, \, n  \; ,   \\
    & X_{2,0} X_{0,2} + X_{2,1} X_{1,2} + X_{23}X_{32} = - m_2 \phi_2 \; ,  \\
    & X_{n-2,n-1} X_{n-1,n-2} + X_{n-2,n} X_{n,n-2} - X_{n-2,n-3}X_{n-3,n-2} = - m_{n-2} \phi_{n-2} \; ,  \\
    & X_{l,l+1}X_{l+1,l} - X_{l,l-1}X_{l-1,l} = - m_l \phi_l \; , \quad l = 3, \, \ldots \, , n-3 \; .
\end{cases}
\end{align}

One can solve the F-terms or employ the graphical computational technique described in \cite{Lindstrom:1999pz}. The detailed computation is given in \cref{app:Dnmoduli}, whereas here we limit ourselves to a summary of the salient points.

First of all, as a consequence of \cref{eq:ftermX}, we have that $\phi_{0,1,n-1,n} = u$, for all external nodes and $\phi_{i=2,\dots,n-2}=$diag$(-u,-u)$, after using the gauge freedom to diagonalize the fields of the internal nodes. Second, from \cref{eq:ftermphi}, dubbing $w_{i,j}=\Tr(X_{i,j}X_{j,i})=w_{j,i}$, we have the following constraints
\begin{align}
    & w_{0,2} = - m_0 u \; , \quad w_{1,2} = - m_1 u \; , \quad  w_{n-1,n-2} = - m_{n-1} u \; , \quad w_{n,n-2} = - m_{n} u \;  \; , \label{eq:zmassesn} \\
    & w_{2,3} = (2 m_2 + m_0 + m_1)u  \; , \label{eq:zmasses2} \\
    & w_{n-2,n-3} = - (2 m_{n-2} + m_{n-1} + m_n)u  \; , \label{eq:zmassesn-2} \\
    & w_{l,l+1} - w_{l,l-1} = 2 m_l u \; , \quad l = 3, \, \ldots \, , n-3 \label{eq:zmassesl} \; .
\end{align}
Finally, by taking the sum of \cref{eq:zmasses2,eq:zmassesn-2,eq:zmassesl}, one then gets the following condition
\begin{align}
    u \left( m_0 + m_1 + \sum_{l=2}^{n-2} 2 m_l + m_{n-1} + m_n \right) = u m = 0 \; ,
\end{align}
while the other gauge invariants constructed out of the chiral fields lead to the algebraic curve describing the moduli space, see \cite{Lindstrom:1999pz}.

As in the $\widehat{A}_{n-1}$ case, we see that there are two possibilities: either the global mass does not vanish, leading to a 2-fold moduli space, or it does and the moduli space is a 3-fold. In the former case, the moduli space is just the Du Val singularity corresponding to $\widehat{D}_n$, while in the latter it is a compound Du Val described by the following equation in $\mathbb{C}^4$
\begin{align}\label{eq:DnFImodulispace}
    x^2 + y^2 w + \beta \, y \, u^n = w^{-1}\left[\prod_{k=1}^n(w + u^2 t_k^2 ) - \prod_{k=1}^n u^2 t_k^2\right] \; ,
\end{align}
where the $t_k$ are given by
\begin{align}
\begin{gathered}
    t_1 = \frac{1}{2} \left( m_0 - m_1 \right) \, ,  \quad
    t_2 = \frac{1}{2} \left( m_{n} - m_{n-1} \right) \, , \quad
    t_3 = \frac{1}{2} \left( m_0 + m_1 \right) \, ,  \\
    t_4 = \frac{1}{2} \left( m_0 + m_1 \right) - m_2 \; , \quad \dots, \quad  t_n = \frac{1}{2} \left( m_0 + m_1 \right) - \sum_{l=2}^{n-2} m_l \; , 
\end{gathered}
\end{align}
and
\begin{align}
    \beta = - 2 \prod_{k=1}^n t_k \; .
\end{align}

Contrary to the previous case, less is known about the existence of local CY metrics on these spaces and thus only the field theory analysis is accessible to study the conformality of the deformed theory. While we leave a detailed analysis to future works, but assuming that the deformed theory flows to an interacting SCFT, we can still analyze the duality group inherited by the deformed theory from the parent one. In particular, we will show that requiring to preserve some duality symmetries of the starting theory, constrains the moduli space of the deformed one.\footnote{One could investigate whether the deformations of Du Val singularities that preserve non-invertible duality defects can be characterized geometrically. Exploring the interplay with deformations that maintain toricity in the $\widehat{A}_{n-1}$ case could lead to deep insights, given the extensive techniques available for studying toric affine Calabi–Yau threefolds and their corresponding $\mathcal N=1$ gauge theories.}

\section{Mass deformations preserving non-invertible symmetries}\label{sec:dualitydefects}

In \cref{sec:globalvariants} we have characterized the locus in the conformal manifolds of $\mathcal N=2$ $\widehat{A}_n$ and $\widehat{D}_n$ quiver SCFTs at which these theories admit non-invertible duality defects: it is the locus which corresponds to symmetric configurations of punctures on the M-theory torus. The study of relevant deformations preserving $\mathcal N=1$ supersymmetry and such non-invertible defects has recently been pioneered in \cite{Damia:2023ses}. Our goal here is to provide a general method to characterize which mass deformations of $\mathcal N=2$ $\widehat{A}_n$ and $\widehat{D}_n$ quivers preserve non-invertible duality, triality or hexality defects. In particular, we derive the dimension of the space of mass deformations which preserve non-invertible defects. In some cases, such deformations lead to known $\mathcal N=1$ SCFTs with moduli spaces Calabi--Yau threefolds, generalizing the deformation of the $\mathcal N=2$ $\widehat{A}_1$ quiver to the conifold SCFT.

The supercharges $Q$ of 4d $\mathcal{N}=4$ theories transform non-trivially under the $\mathrm{SL}(2,\mathbb{Z})$ duality group \cite{Kapustin:2006pk}. More precisely, they transform as $Q \to \exp(-i \beta) Q$. At self-dual values of the coupling $\tau$, it turns out that $\beta = \pi i / q$, where $q$ is the order of the stabilizer of $\tau$ in $\mathrm{SL}(2,\mathbb{Z})$. In $\mathcal N=1$ superspace, a mass deformation takes the form
\begin{align}
    \Delta S_W = \int \text{d}^2 \theta~ \frac{m}{2} \Phi_i^2 \, .
\end{align}
Such a deformation preserves the duality symmetry if and only if the transformation of the measure $\text{d}^2 \theta\to \exp(-2i \beta)\text{d}^2 \theta $ can be reabsorbed by a transformation of the chiral fields $\Phi_i$. This can be done using the R-symmetry \cite{Damia:2023ses}, if the masses $(m_1,\dots,m_n)$ defining the deformation satisfy
\begin{equation}
    (m_1',\dots,m_n') = e^{i\alpha} (m_1,\dots,m_n)~,
\end{equation}
where $(m_1',\dots,m_n')$ is the image of $(m_1,\dots,m_n)$ under the duality transformation at hand.

We have seen that dualities sometimes act non trivially on the deformation masses, which makes the analysis more involved. This happens when the global mass does not vanish; therefore, in what follows we distinguish the cases with vanishing global mass with the ones where the global mass is non-zero.

\subsection{Vanishing global mass}\label{Sec:vanishingglobmass}

When the global mass $m$ vanishes, the action of $\mathrm{SL}(2,\mathbb{Z})$ and deck transformations on the $z_i$'s is trivial. This means that, while the punctures $p_i$ are rotated to $p_i'$, cf. \cref{sec:Andualitysymm}, the image $p_i'$ of each $p_i$ is still associated with the original $z_i$. Consider a configuration of $n$ punctures invariant under a duality transformation of order $q$, as discussed in \cref{sec:Andualitysymm}, schematically $\mathcal{D} = \sigma \circ t \circ S/ST$ is a composition of permutation, deck and $\mathrm{SL}(2,\mathbb{Z})$ operations. $\mathcal{D}$ acts trivially on $\vec{p}$ by construction, while only $\sigma$ acts on $\vec{z}$ 
\begin{equation}
    \mathcal{D}\left((p_1,z_1),(p_2,z_2),\dots,(p_n,z_n)\right) = \left((p_1,z_{\sigma(1)}),(p_2,z_{\sigma(2)}),\dots,(p_n,z_{\sigma(n)})\right)~.
\end{equation}
Therefore, for the masses to preserve the duality symmetry, they must solve the eigenvalue problem
\begin{align}\label{eq:eigenprobl}
    \mathcal{D} \, \vec{m} = e^{i \alpha} \, \vec{m} \; ,
\end{align}
where $\mathcal{D}$ acts on the masses only through the subgroup of permutations in the whole duality group. The $n$ `decorated' punctures $(p_i,z_i)$ split into orbits under $\mathcal D$ of size a divisor of the order of $\mathcal D$. For example, under $\mathcal{D} = \sigma \circ t \circ S$ at $\tau=i$, the punctures always split into $O_4$ orbits of size 4, $O_2\leq 1$ orbits of size 2 and $O_1\leq 2$ orbits of size 1,\footnote{Note that some configurations require that one or more gauge groups have infinite coupling even if punctures are distinct, for example when $O_2=O_1=1$.} with a total number of distinct orbits $O_{tot} = O_1 + O_2 + O_4$. 

The total number of solutions to \cref{eq:eigenprobl} can be explicitly given in terms of the number of orbits. Let $q$ be the order of $\mathcal{D}$, thus $\mathcal{D}^q=1$ implies that the phase $e^{i \alpha}$ is a $q^{th}$ root of unity. Then, the total number of independent mass deformations satisfying \cref{eq:eigenprobl} is given by
\begin{itemize}
    \item If $e^{i\alpha}\neq 1$, there is one deformation for each orbit of size $k$ such that $e^{ik\alpha}=1$. Hence the total of independent deformations is $\sum_k O_k$.
    
    \item If $e^{i\alpha}=1$, then the number of independent deformations is $O_{tot}-1$.
\end{itemize}
We refer the reader to \cref{app:generalproof} for a complete and detailed proof of this result. 

The space of solutions represent all of the possible relevant deformations of the $\mathcal{N}=2$ theory that preserve the non-invertible duality symmetry, and distinct solutions flow in principle to distinct $\mathcal{N}=1$ SCFTs. The moduli space of the IR theories is then specified by each mass deformation, as discussed in \cref{sec:modulispace}. 

We dedicate the rest of this section to explicit examples.

\paragraph{Mass deformed \texorpdfstring{$\widehat{A}_3$}{A3} quiver theory}

\begin{figure}[h!]
    \centering
    \includegraphics[]{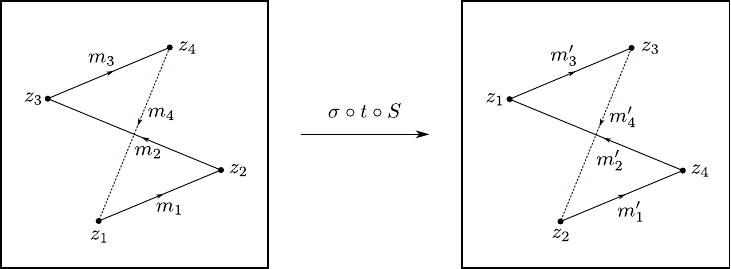}
    \caption{Action of the defect $\mathcal{D}= \sigma\circ t\circ S$ for $\widehat{A}_3$.}\label{fig:Sselfdualconfigmasses}
\end{figure}

We consider the configuration shown in \cref{fig:Sselfdualconfigmasses}. At $\tau=i$, $\mathcal{D} = \sigma \circ t \circ S$ defines a non-invertible duality defect, where $t=\prod_i t^{\tau}_i$ and $\sigma= s_2 s_1 s_3$. If one then turns on mass deformations in such a way that the global mass vanishes, $\mathcal D$ acts on the masses as
\begin{align}
    (m_1',m_2',m_3',m_4') &=(z_2'-z_1',z_3'-z_2',z_4'-z_3',z_1'-z_4') \nonumber \\
    &=(z_4-z_2,z_1-z_4,z_3-z_1,z_2-z_3) \nonumber \\
    &=(m_2+m_3,m_4,m_1+m_2,-m_2)~.
\end{align}

The condition on the masses to preserve the non-invertible duality defect reads
\begin{equation}
    (m_1',m_2',m_3',m_4')=e^{i \alpha}(m_1,m_2,m_3,m_4)~,
\end{equation}
with solutions:
    \begin{align}
        &(m_1, \dots, m_4) = \left(1,0,-1,0\right) m_1 \quad \text{for } \alpha=\pi \, , \nonumber \\[4pt]
        &(m_1, \dots, m_4) = \left(\frac{-i-1}{2},1,\frac{-i-1}{2},i \right) m_2 \quad \text{for } \alpha=\frac{\pi}{2} \, , \nonumber \\[4pt]
        &(m_1, \dots, m_4) = \left(\frac{i-1}{2},1,\frac{i-1}{2},-i \right) m_2  \quad \text{for } \alpha=-\frac{\pi}{2} \, .
    \end{align}
In other words, inside the (complex) 3-dimensional space of mass deformations for the $\mathcal N=2$ $\widehat{A}_3$ SCFT, the subspace of deformations preserving non-invertible duality defects is one-dimensional. More precisely, it is the union of three lines.

As discussed in \cref{sec:modspaceAn}, these three mass eigenvectors determine the moduli space of the IR SCFT at the end of the RG flow. For $\alpha=\pi$ we have
\begin{align}
    xy = w^2 \left( w - m_1 u \right)^2 \; ,
\end{align}
which is the equation of the toric $L^{2,2,2}$ singularity. In contrast, for $\alpha=\pm\frac{\pi}{2}$ the moduli space is defined by
\begin{align}
    xy = w^4 - \left( m_2 \frac{u}{2} \right)^4 \; .
\end{align}

\begin{figure}[h!]
    \centering
    \includegraphics[]{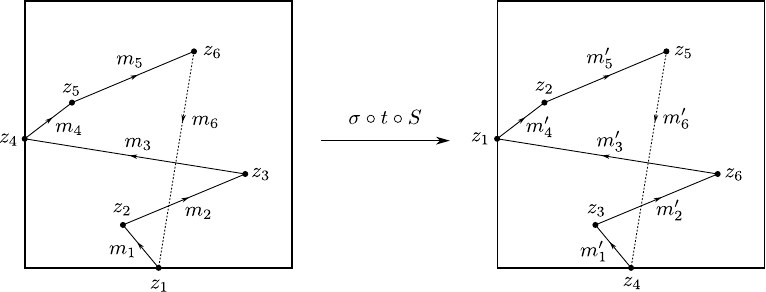}
    \caption{Action of the defect $\mathcal{D}= \sigma\circ t\circ S$ for $\widehat{A}_5$.}\label{fig:SselfdualconfigSection6}
\end{figure}

\paragraph{Mass deformed $\widehat{A}_5$ quiver theory} We consider the $S$-self dual configuration of 6 punctures on the M-theory torus $E_i$ shown in \cref{fig:SselfdualconfigSection6}. Note that the six punctures split into a generic orbit of size four $(2~5~6~3)$, and a non-generic one of size two $(1~4)$. The combination $\mathcal{D}=\sigma \circ t \circ S$, where 
\begin{equation}
    t=\prod_{i=1}^6 t_i^{(\tau)}~~~\text{and}~~~\sigma = s_3 s_2 s_1 s_4 s_3 s_2 s_3 s_5~,
\end{equation}
acts on the masses as 
\begin{align}
    (m_1,m_2,m_3,m_4,m_5,m_6) \to (-m_3, m_3 + m_4 + m_5, m_6, m_1, m_2 + m_3 + m_4, -m_4) \, .
\end{align}

Solving the eigenvalue equation $\mathcal{D} \vec{m} = e^{i\alpha} \vec{m}$ leads to the following five solutions
\begin{equation}\label{eq:massphases6BB}
    \begin{array}{lll}
        \alpha=\pi & \left(1,0,1,-1,0,-1\right) m_1 & xy = w^4 \left[ w^2 - (u m_1)^2\right] \\[4pt]
        \alpha=\pi & \left(0,1,0,0,-1,0 \right) m_2 & xy = w^3 \left( w - u m_2\right)^3 \\[4pt]
        \alpha=\frac{\pi}{2} \hspace{0.5cm} & \left(1,i-1,-i,-i,i-1,1 \right) m_1 \hspace{0.5cm} & xy = w^2 \left[ w^4 - (u m_1)^4\right] \\[4pt]
        \alpha=-\frac{\pi}{2} & \left(1,-i-1,i,i,-i-1,1 \right) m_1 & xy = w^2 \left[ w^4 - (u m_1)^4\right] \\[4pt]
        \alpha=0 & \left(1,0,-1,1,0,-1\right) m_1 & xy = w^2 \left( w - u m_1\right)^4 \; .
    \end{array}
\end{equation}
In accordance with the general analysis, we find that there are five independent mass deformation, where only two of them can be turned on simultaneously, i.e.~the ones associated to $\alpha=\pi$. The moduli space of the second and last solutions are toric singularities usually referred to as $L^{3,3,3}$ and $L^{2,4,2}$, respectively. 

We give more examples and details in \cref{app:examples}.

\begin{figure}[h!]
    \centering
    \includegraphics[]{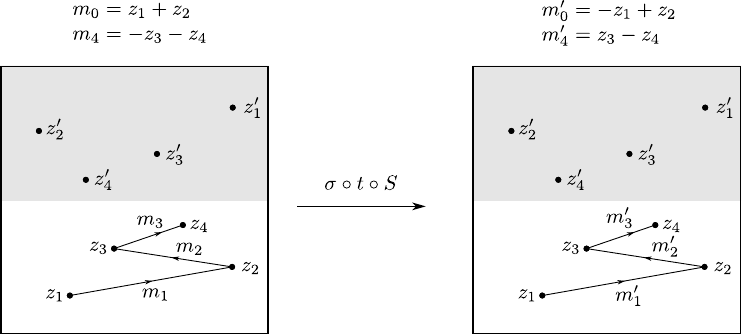}
    \caption{Action of the defect $\mathcal{D}= R_{D,4} \circ R_{D,2} \circ \sigma\circ t\circ S$ for $\widehat{D}_4$.}\label{fig:SselfdualconfigtypeD}
\end{figure}

\paragraph{Mass deformed $\widehat{D}_4$ quiver theory} The same analysis applies to the $\widehat{D}_n$ case. For instance, let us consider $\widehat{D}_4$, which requires 8 marked points to be placed on the torus $\mathbb{T}^2$, and organize them in two orbits of size four. In \cref{sec:Andualitysymm}, we showed that this configuration leads to a theory with a duality defect $\mathcal{D} = R_{D,4} \circ R_{D,2} \circ \sigma \circ t^{(i)} \circ S$, with $\sigma = s_1 s_3$ and\footnote{Recall that for $\widehat{D}_n$, $t_{k}^{(\tau)} = R_{C,k} \circ R_{A,k}$, acting on both marked points and images.} $t^{(i)} = \prod_{k=1}^4 t_k^{(i)}$. The action of this defect is depicted in \cref{fig:SselfdualconfigtypeD}. Under $\mathcal{D}$, the masses transform as 
\begin{align}
        \mathcal{D} : (m_0,m_1,m_2,m_3,m_4) \to (m_1, -m_0, m_0 + m_2 + m_3, m_4, -m_3) \, .
\end{align}
The solutions of the eigenvalues equation $\mathcal{D} \vec{m} = e^{i \alpha} \vec{m}$ for vanishing global mass are 
\begin{equation}
    \begin{array}{lll}
        \alpha = \frac{\pi}{2} \quad & (i,-1,0,-i,1) m_4 & x^2 + y^2 w = w^3 + \frac{1}{2} \left(m_4 u \right)^4 (y+w)  \\[4pt]
        \alpha = \frac{\pi}{2} & (i-1,-i-1,1,0,0) m_2 \quad & x^2 + y^2 w = \left[ w^2 - (m_2 u)^4 \right] \left( w + 4 m_2 u \right) \\[4pt]
        \alpha = -\frac{\pi}{2} & (-i,-1,0,i,1) m_4 & x^2 + y^2 w = w^3 + \frac{1}{2} \left(m_4 u \right)^4 (y+w) \\[4pt]
        \alpha = -\frac{\pi}{2} & (-i-1,i-1,1,0,0) m_2 & x^2 + y^2 w = \left[ w^2 - (m_2 u)^4 \right] \left( w + 4 m_2 u \right) \; . 
    \end{array}
\end{equation}
    
\subsection{Non-vanishing global mass}

The case of non-vanishing global mass is more involved, since now $\mathcal{D}$ acts non-trivially on the $z_i$. However, the punchline is the same. One applies the transformation $\mathcal{D}$ to the masses using \cref{eq:Anmasses} and looks for an eigenvector of masses with the further constraints that the global mass change as $m' =  m/\tau$, which forces the phase to be $e^{i\alpha}=1/\tau$. 

We can again prove in this case that a solution to the eigenvalue problem $\mathcal{D} \vec{m} = e^{i \alpha} \vec{m}$ always exists. To this end, we need to prove that there is at least one eigenvector with eigenvalue $1/\tau$, as required by the transformation properties of the global mass. One can check explicitly that in this case the Dynkin vector $\vec{n}$ is a left eigenvector with eigenvalue $1/\tau$.\footnote{The vector $\vec{n}$ is a left eigenvector with eigenvalue $1$ for both the $t_i^{\tau}/R_{I,i}$ and $s_i$ transformations, and it has eigenvalue $1/\tau$ for the modular transformation $S$.} Since right and left eigenvectors of an automorphism form a basis for a vector space and its dual respectively, we have that at least one right eigenvector $\vec{m}$ exists, with eigenvalue $1/\tau$, such that $m = \vec{n} \cdot \vec{m} \neq 0$. This proves there is always a mass deformation, with non-vanishing global mass, that preserves the duality symmetry. Moreover, for each other right eigenvector with eigenvalue $1/\tau$, we have an extra dimension in the space of solutions to \cref{eq:eigenprobl}.\footnote{Contrary to the case of non-vanishing global mass, we do not have a general formula for the dimension of this space.}

As an example, consider $\widehat{A}_3$ and the configuration of \cref{fig:Sselfdualconfigmasses}, but this time in the presence of a global mass. In order to the duality defect $\mathcal{D} = \sigma \circ t \circ S$, the masses need to transform as
    \begin{align}
        m_2 + m_3 - m \tau_1 &= -i m_1 \; , \nonumber \\
        m_4 - m \left[ \tau_1(i-1) + 1 \right] &= - i m_2  \; , \nonumber \\
        m_1 + m_2 - m \tau_1 &= -i m_3 \; , \nonumber \\
        -m_2 + m \left[ \tau_1 (i+1)-i \right] &= -i m_4 \; ,
    \end{align}
and there are two independent mass deformations with $\mathcal{D}$:
    \begin{align}
        &\left( m_1 , m_2, m_3, m_4  \right) = \left( 0 , \frac{\tau_1}{1 - \tau_1}, 0, 1 \right) m_4 \; , \qquad m = \frac{m_4}{1-\tau_1} \; , \nonumber \\[4pt]
        &\left( m_1 , m_2, m_3, m_4  \right) = \left( 1, \frac{2 \tau_1 - i + 1}{1 - \tau_1}, 1, 0 \right) m_1 \; , \qquad m = m_1 \frac{1-i}{1-\tau_1} \; , 
    \end{align}
where $\alpha=-\pi/2$. The moduli space of these IR SCFTs is a 2-fold, as discussed in \cref{sec:modulispace}.

In the mass deformed $\widehat{A}_5$ theory with global mass, the condition to preserve $\mathcal{D}$ is
    \begin{align}
        -m_3 & = -i m_1 + m (\tau_1 - i) \; , \nonumber \\
        m_3 + m_4 + m_5 & = - i m_2 + m \left[ \tau_1 (i-1) + \frac{i+1}{2} \right] \; , \nonumber \\
        m_6 & = -i m_3 - m i \tau_1 \; , \nonumber \\
        m_1 & = -i m_4 - m i \tau_1 \; , \nonumber \\
        m_2 + m_3 + m_4 & = -i m_5 + m \left[ \tau_1 (i-1) + \frac{i+1}{2} \right] \; , \nonumber \\
        - m_4 & = - i m_6 + m \left( \tau_1 + i \right) \; ,
    \end{align}
    which has two eigenvectors with eigenvalue $\alpha=-\pi/2$ 
    \begin{align}
        &\left( \frac{2\tau_1 -i}{2 \tau_1 + i +2} , \, 0 , \, \frac{-2\tau_1 + 2i - 1}{2 \tau_1 + i +2} , \, \frac{1 - 2\tau_1}{2 \tau_1 + i +2} , \, 0, \, 1 \right) m_6 \; , \qquad m = \frac{2 + 2i}{2 \tau_1 + i +2}m_6 \nonumber \\[4pt]
        &\left( \frac{-2}{2 \tau_1 + i +2} , \, 1 , \, \frac{-2\tau_1 }{2 \tau_1 + i +2} , \, \frac{- 2\tau_1 - 2i}{2 \tau_1 + i +2} , \, 1, \, 0 \right) m_2 \; , \qquad m = \frac{2 + 2i}{2 \tau_1 + i +2}m_2
    \end{align}
    and in both cases the IR theory has moduli space given by $\mathbb{C}^2/\mathbb{Z}_5$, from the discussion in \cref{sec:modspaceAn}.

As a $\widehat{D}_n$ example, consider the case with $n=4$ and two orbits of size four, for which we find that there are two mass configurations that preserve $\mathcal{D} = R_{D,4} \circ R_{D,2} \circ \sigma \circ t^{(i)} \circ S$ with $\alpha = \pi/2$
\begin{align}
     &\left( m_0 , \, \ldots \, , m_4 \right) = \left(-\frac{i+1}{2}\left( \frac{\tau_0 + \tau_3 - 2}{\tau_3} \right), \frac{i+1}{2}\left( \frac{\tau_0 + i \tau_3 - i - 1}{\tau_3} \right), 0, 0, 1 \right) m_4 \; , \nonumber \\[5pt]
     &\left( m_0 , \, \ldots \, , m_4 \right) = \left(\frac{-i+1}{2}\left( \frac{\tau_0 - i \tau_3 - 2}{\tau_3} \right), - \frac{-i+1}{2}\left( \frac{\tau_0 + \tau_3 - i - 1}{\tau_3} \right), 0, 1, 0 \right) m_3 \; , 
\end{align}
with global mass $m=m_4/\tau_3$ and $m=i m_3/\tau_3$ respectively. From the discussion in \cref{sec:modspaceDn}, in both cases the moduli space is simply the 2-fold Du Val singularity of type $D$.

More examples can be found in \cref{app:examples}.

\section*{Acknowledgements}

The authors would like to thank Simone Giacomelli, Azeem Hasan, Elias Riedel Gårding and Luigi Tizzano for useful comments and clarifying discussions. 
R.A. and A.C. are respectively a Research Director and a Senior Research Associate of the F.R.S.-FNRS (Belgium). 
The work of S.M. is supported by ``Fondazione Angelo Della Riccia'' and by funds from the Solvay Family. S.N.M. acknowledges the support from the Simons Foundation (grant \#888984, Simons Collaboration on Global Categorical Symmetries) as well as the European Research Council (ERC) under the European Union’s Horizon 2020 research and innovation program (grant agreement No. 851931). V.T. is funded by the Deutsche Forschungsgemeinschaft (DFG, German Research Foundation) under Germany’s Excellence Strategy EXC 2181/1 - 390900948 (the Heidelberg STRUCTURES Excellence Cluster). This research is further supported by IISN-Belgium (convention 4.4503.15) and through an ARC advanced project.

\appendix

\section{Details on the F-terms of mass deformed \texorpdfstring{$\widehat D_n$}{Dn}}\label{app:Dnmoduli}

In this section we want to identify the moduli space of the $\mathcal{N}=1$ theory obtained by mass deformation of the $\widehat{D}_n$-shaped quiver gauge theory with gauge group $\mathrm{SU}(k)^4 \times \mathrm{SU}(2k)^{n-3}$. 

We will use the following convention for fields: denote the adjoint fields with $\phi_i$, the chiral fields that transform in the fundamental representation of an external node of the quiver with $A$ and in their anti-fundamental representation with $B$, while the remaining field transforming in the bifundamental representation of $\mathrm{SU}(2k)$ gauge factors with $X_{i,i+1}$ so that
\begin{align}
    & A_i=\left( \tiny{\yng(1)}_i, \overline{\tiny{\yng(1)}}_2 \right) \; , \quad i = 0,\, 1 \; , \nonumber \\
    & B_i=\left( \tiny{\yng(1)}_2, \overline{\tiny{\yng(1)}}_i \right) \; , \quad i = 0,\, 1 \; , \nonumber \\
    & A_j=\left( \tiny{\yng(1)}_j, \overline{\tiny{\yng(1)}}_{n-2} \right) \; , \quad j = n-1, \, n \; , \nonumber \\
    & B_j=\left( \tiny{\yng(1)}_{n-2}, \overline{\tiny{\yng(1)}}_j \right) \; , \quad j = n-1, \, n \; , \nonumber \\
    & X_{l,l+1}=\left( \tiny{\yng(1)}_l , \overline{\tiny{\yng(1)}}_{l+1} \right) \; , \quad l = 2, \, \ldots \, , n-3 \; ,
\end{align}
and accordingly for $X_{l+1,l}$. The generic superpotential deformed with mass terms for adjoints reads
\begin{align}
    W_{\mathcal{N}=1} &= \sum_{i=0,1} \phi_i B_i A_i + \sum_{j=n-1,n} \phi_{j} B_{j} A_{j} \nonumber \\
    &+ \phi_2 \left( A_0 B_0 + A_1 B_1 + X_{23} X_{32} \right) + \phi_{n-2} \left( A_{n-1} B_{n-1} + A_{n} B_{n} - X_{n-2,n-3} X_{n-3,n-2} \right) \nonumber \\
    &+ \sum_{l=3}^{n-3} \phi_l \left( X_{l,l+1}X_{l+1,l} - X_{l,l-1}X_{l-1,l} \right) + \sum_{i=0}^{n} \frac{m_i}{2} \phi_i^2 \; .
\end{align}
We need to solve the F-terms, in order to find the equation that defines the moduli space and how it is affected by the choice of the masses. For the first goal, we rely on the computation carried out in \cite{Lindstrom:1999pz} via the bug calculus graphical approach. In the following, we explicitly show how the masses affect the value of $\Tr \, \phi_i$. The F-terms for the chiral fields are
\begin{align}
    & \phi_2 A_i = - A_i \phi_i \; , \quad i = 0, \, 1  \; , \label{eq:eigenvalueA0} \\
    & B_i \phi_2 = - \phi_i B_i \; , \quad i = 0, \, 1  \; , \label{eq:eigenvalueB0} \\
    & \phi_{n-2} A_j = - A_j \phi_{j} \; , \quad j = n-1, \, n \; , \label{eq:eigenvalueAn} \\
    & B_j \phi_{n-2} = - \phi_{j} B_j \; , \quad j = n-1, \, n  \; , \label{eq:eigenvalueBn} \\
    & \phi_l X_{l,l+1} = X_{l,l+1} \phi_{l+1} \; , \quad l = 2, \, \ldots \, , n-2 \; , \label{eq:eigenvalueX} \\
    & X_{l+1,l} \phi_{l} = \phi_{l+1} X_{l+1,l} \; , \quad l = 2, \, \ldots \, , n-2 \; , \label{eq:eigenvalueX+1} \; ,
\end{align}
while for the adjoint fields 
\begin{align}
    & B_i A_i = - m_i \phi_i \; , \quad i = 0, \, 1  \; , \label{eq:eigenmasses0} \\
    & B_j A_j = - m_j \phi_j \; , \quad j = n-1, \, n  \; , \label{eq:eigenmassesn} \\
    & A_0 B_0 + A_1 B_1 + X_{23}X_{32} = - m_2 \phi_2 \; , \label{eq:eigenmasses2} \\
    & A_{n-1} B_{n-1} + A_n B_n - X_{n-2,n-3}X_{n-3,n-2} = - m_{n-2} \phi_{n-2} \; , \label{eq:eigenmassesn-2} \\
    & X_{l,l+1}X_{l+1,l} - X_{l,l-1}X_{l-1,l} = - m_l \phi_l \; , \quad l = 3, \, \ldots \, , n-3 \label{eq:eigenmassesl} \; .
\end{align}
As we did for the $\widehat{A}_{n-1}$, we consider the moduli space of the theory with gauge group $\mathrm{U}(1)^4 \times \mathrm{U}(2)^{n-3}$, and the generic case will be given by the $k$-th symmetric product of this space. 
Let us proceed in steps. First, we show that $-\phi_0$ and $-\phi_1$, which are complex numbers, are the eigenvalues of $\phi_2$, which is a $2 \times 2$ matrix. \Cref{eq:eigenvalueA0} and \cref{eq:eigenvalueB0} have the form of a right and left eigenvalue equation for $\phi_2$, and there are two eigenvectors $A_0$, $B_0^T$ with eigenvalue $-\phi_0$, and two $A_1$, $B_1^T$ with eigenvalue $-\phi_1$. Assume for now that none of them is a null vector, otherwise either $m_i=0$ or $\phi_i=0$. By \cref{eq:eigenmasses0} we see that $A_i$ and $B_i$ are not orthogonal, and there is no relation between $A_0$ and $A_1$. The way to accommodate them is that $B_0^T$ and $B_1^T$ are proportional to $A_0$ and $A_1$, respectively, and the latter are linearly independent. So it exists a matrix $V_2$, whose columns are the two eigenvectors $A_0$ and $A_1$, that diagonalizes $\phi_2$. On the other hand, if we lift the non-null assumption, and consider that, say, $B_0 = 0 = B_1$, and $m_0 = m_1 = 0 $ and again we are left with $A_0$ and $A_1$ as the two eigenvectors of $\phi_2$. A similar reasoning holds for $\phi_{n-2}$, whose eigenvalues are $-\phi_{n-1}$ and $-\phi_n$ with eigenvectors $A_{n-1}$ and $A_n$. Hence
\begin{align}
    &\phi_2^d = V_2^{-1} \phi_2 V_2 = \left( 
    \begin{array}{cc}
       - \phi_0  & 0 \\
        0 & - \phi_1
    \end{array}
    \right) \; \nonumber \\[5pt]
    &\phi_{n-2}^d = V_{n-2}^{-1} \phi_{n-2} V_{n-2} = \left( 
    \begin{array}{cc}
       - \phi_{n-1}  & 0 \\
        0 & - \phi_n
    \end{array}
    \right) \; . 
\end{align}
As a second step, we show that all of these eigenvalues are equal. Consider
\begin{align}
    B_i \phi_2 X_{23} X_{34} \ldots X_{l,l+1} \ldots X_{n-3,n-2} A_j \; , \quad i=0,1 \; , \: j=n-1,n \; ,
\end{align}
and by \cref{eq:eigenvalueB0} and \cref{eq:eigenvalueX} we can write it in two equivalent ways
\begin{align}\label{eq:secondstepPhi}
    - \phi_i B_i X_{23} X_{34} \ldots X_{l,l+1} \ldots X_{n-3,n-2} A_j = B_i X_{23} \phi_3 X_{34} \ldots X_{l,l+1} \ldots X_{n-3,n-2} A_j \; , \quad &i=0,1 \; , \nonumber\\
    &j=n-1,n \; .
\end{align}
Using recursively \cref{eq:eigenvalueX} and \cref{eq:eigenvalueX+1} we can move the adjoint until the end 
\begin{align}
    B_i X_{23} X_{34} \ldots X_{l,l+1} \ldots X_{n-3,n-2} \phi_{n-2} A_j \; , \quad i=0,1 \; , \: j=n-1,n \; ,
\end{align}
where we can use \cref{eq:eigenvalueAn} to write 
\begin{align}
    -B_i X_{23} X_{34} \ldots X_{l,l+1} \ldots X_{n-3,n-2} A_j \phi_j \; , \quad i=0,1 \; , \: j=n-1,n \; ,
\end{align}
and comparing with \cref{eq:secondstepPhi} we obtain
\begin{align}\label{eq:allphiequal}
    &\phi_i = \phi_j := u \; , \quad i=0,1 \; , \: j=n-1,n \; , \nonumber \\
    &\phi_2^d = \phi_{n-2}^d = - u \, {1}_{2\times 2} \; .
\end{align}

As a third step, we show that all 2-dimensional matrices have the same eigenvalues. From \cref{eq:eigenvalueX}, consider $l=2$, diagonalize $\phi_2$ and use \cref{eq:allphiequal}
\begin{align}
    \phi_2 X_{23} = X_{23} \phi_3 = V_2 V_2^{-1} \phi_2 V_2 V_2^{-1}X_{23} = V_2 \, \phi_2^d \, V_2^{-1} X_{23} = -u V_2 {1}_{2\times2} V_2^{-1} X_{23} = - u X_{23} \; ,
\end{align}
and from second and last step this is now a left eigenvalue equation for $\phi_3$, with eigenvalue $\phi$ associated to $X_{23}$. The same reasoning can be repeated for $X_{32}\phi_2 = \phi_3 X_{32}$ from \cref{eq:eigenvalueX+1}, obtaining that $X_{23}$ and $X_{32}$ are the eigenvectors of $\phi_3$ with eigenvalue $-u$. We get that $\phi_2^d = \phi_3^d$. We can recursively repeat the argument for all $l$, obtaining
\begin{align}
    \phi_l = - u \, {1}_{2 \times 2} \; , \quad l = 2, \, \ldots \, n-2 \; .
\end{align}
Note that the same reasoning holds in case of all vanishing masses, so that $u$ is the variable that parametrizes the $\mathbb{C}$ factor in the moduli space of the $\mathcal{N}=2$ theory.

As a fourth step, we construct the $\Tr \, X_{l,l+1}X_{l+1,l}$. Taking the trace of \cref{eq:eigenmasses2} and \cref{eq:eigenmassesn-2}, and using \cref{eq:eigenmasses0}-\cref{eq:eigenmassesn} and the fact that $\Tr \, \phi_2 = \Tr \, \phi_{n-2} = - 2 u$, we get
\begin{align}
    &\Tr \, X_{23}X_{32} = u \left( m_0 + m_1 + 2m_2 \right) \; , \label{eq:tracesmasses23} \\
    -&\Tr \, X_{n-2,n-3}X_{n-3,n-2} = u \left( m_{n-1} + m_n + 2m_{n-2} \right) \; . \label{eq:tracesmassesn-2n-3}
\end{align}
Similarly, from \cref{eq:eigenvalueX} we find that
\begin{align}
    \Tr \, X_{l,l+1}X_{l+1,l} = \Tr \, X_{l,l-1}X_{l-1,l} - m_l \Tr \, \phi_l = \Tr \, X_{l,l-1}X_{l-1,l} + 2 m_l u \; , \quad l = 3, \, \ldots \, , n-3 \; ,
\end{align}
and using it recursively we get that 
\begin{align}\label{eq:tracesrecursionmasses}
    \Tr \, X_{n-3,n-2}X_{n-2,n-3} = \Tr \, X_{23}X_{32} + u \sum_{l=3}^{n-3} 2 m_l \; .
\end{align}
By inserting \cref{eq:tracesrecursionmasses} in \cref{eq:tracesmassesn-2n-3} and summing with \cref{eq:tracesmasses23} we obtain
\begin{align}
    u \left( m_0 + m_1 + \sum_{l=2}^{n-2} 2 m_l + m_{n-1} + m_n \right) = u m = 0 \; .
\end{align}
Similarly to what happens in the $\widehat{A}_{n-1}$ case, the global mass and the value of the adjoint fields are related: when the global mass is zero, $u$ can be non-zero, while when the global mass $m\neq 0$, it is forced $u=0$. 

Finding the form of moduli space for $m=0$ and $u\neq0$ by solving directly the F-terms is quite involved. As for $\widehat{A}_{n-1}$, in \cite{Lindstrom:1999pz} they deform the quiver gauge theory by FI terms $b_i$ at each node and they carry out this computation exploiting the graphical tool of bug calculus. By comparison of the F-terms in \cref{eq:eigenvalueA0} : \cref{eq:eigenvalueX+1} with the graphical representation in \cite{Lindstrom:1999pz}, we can identify
\begin{align} \label{eq:FImassesDn}
    b_i \leftrightarrow - m_i \, \phi_i \;, \quad \forall i \; ,
\end{align}
where all FI-terms are subject to the condition 
\begin{align}
     b_0 + b_1 + b_{n-1} + b_n = 2 \sum_{i=2}^{n-2} b_i \; ,
\end{align}
which translates in the trace of the sum of $m_i \phi_i$, i.e.  
\begin{align}
    \left( m_0 + m_1 + \sum_{l=2}^{n-2} 2 m_l + m_{n-1} + m_n \right) u = 0 \; .
\end{align}

\section{On \texorpdfstring{$\mathcal N=1$}{N=1} mass deformations preserving non-invertible symmetries}\label{app:generalproof}

We systematically study the solutions to the eigenvalue problem
\begin{align}\label{eq:eigenapp}
    \mathcal{D} \, \vec{m} = e^{i \alpha} \, \vec{m} 
\end{align}
when the global mass vanishes. We consider the action of the permutation first on the $z_i$, and then on the masses $m_i$. Let $\mathcal Z$ be the vector space spanned by the $z_i$'s in $\widehat{A}_{n-1}$ configurations. 

Note that $\mathcal{D}$ can be block diagonalized in $\mathcal Z$ according to the orbit decomposition discussed in \cref{sec:Andualitysymm}. Each block is a finite order matrix and hence can be diagonalized. Moreover, the minimal polynomial of each block is $x^n-1$, where $n$ is the order of the orbit, and it divides the characteristic polynomial of the block, which is of the same order. Therefore, each orbit contributes $n$ eigenvalues, specifically $n$ districts roots of unity. Now, the masses $m_i$ span a codimension one subspace $\mathcal M$ of $\mathcal Z$. The direction orthogonal to $\mathcal M$ in $\mathcal Z$ is generated by the vector of Dynkin labels $\vec{n}$ since $\vec{m} \cdot \vec{n} = 0$, and it is associated with an eigenvector of the permutation matrix with eigenvalue $1$. Thus, the defect $\mathcal{D}$ acting on $\mathcal M$ retains all the eigenvectors but the one dual to $\vec{n}$. 

In conclusion, within the space of mass deformations solving \cref{eq:eigenapp}, those corresponding to the eigenvalue $e^{i \alpha}$ where $(e^{i \alpha})^k=1$ and $e^{i \alpha}\neq 1$, span a subspace of dimension the number of orbits of order a multiple of $k$. Mass deformations solving \cref{eq:eigenapp} with eigenvalue $e^{i \alpha}=1$ rather span a subspace of dimension the total number of orbits minus 1. This reasoning applies both for duality and triality symmetries of $\widehat{A}_{n-1}$ quivers. Since every $\widehat{D}_n$ configuration of marked points can be seen as special $\widehat{A}_{2n}$ configuration, where the tiltings associated to a puncture and its image under $R_D$ satisfy $z_{i'}=-z_i$, the logic also applies to $\widehat{D}_n$ quivers provided one restricts to the subspace of $\mathcal Z$ (and $\mathcal M$) satisfying this additional constraint.

This rationale translates into an efficient method for computing which mass deformations preserve non-invertible duality defects in general cases. While the system of \cref{eq:eigenapp} can in principle always be explicitly solved by brute force, in practice it becomes rapidly cumbersome as the number of punctures grows. However, the underlying orbit structure allows the advertised more efficient calculation.

The main point is to trade the `physical' basis of $\mathcal M$ for another basis adapted to the orbit decomposition under $\mathcal D$. For example, in $\widehat{A}_3$ as studied in \cref{Sec:vanishingglobmass}, a convenient choice\footnote{Since $n_1+n_2+n_3+n_4=0$, here ``basis" is to be understood as ``generating set".} is:
\begin{equation}
    (n_1,n_2,n_3,n_4)=(z_3-z_1,z_4-z_3,z_2-z_4,z_1-z_2)~.
\end{equation}
It satisfies the appreciable property that
\begin{equation}
    (n_1',n_2',n_3',n_4')=(n_4,n_1,n_2,n_3)~,
\end{equation}
which in turn simplifies the analysis of the condition $(n_1',n_2',n_3',n_4') =\alpha(n_1,n_2,n_3,n_4)$: $e^{i \alpha}$ must satisfy $e^{4 i \alpha}=1$ and $e^{i \alpha} \neq 1$ (so that the global mass vanishes). This result is equivalent to the one of \cref{Sec:vanishingglobmass}, as
\begin{equation}
    n_1 = m_1+m_2~,~~~n_2 = m_3~,~~~n_3=-m_2-m_3~,~~~n_4 = -m_1~,
\end{equation}
is invertible. 
    
\begin{figure}[h!]
    \centering
    \includegraphics[]{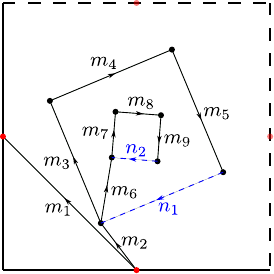}
    \caption{General $t\circ S$-symmetric configuration of punctures in $E_i$. }\label{fig:Sselfdualconfigtree}
    \end{figure}

To analyze a general configuration one needs to group the punctures into orbits under $t\circ S$. An example is shown in \Cref{fig:Sselfdualconfigtree}, which displays a configuration consisting of two orbits of size four and one orbit of size two under $\sigma\circ t\circ S$ on $E_i$.
    
The 9-uplet $(m_1,\dots,m_9)$ is a basis of $\mathcal{M}$ and we also show $n_1=-(m_3+m_4+m_5)$ and $n_2=-(m_7+m_8+m_9)$ in \Cref{fig:Sselfdualconfigtree} for symmetry. Under $\sigma\circ t\circ S$:
\begin{equation}
    (m_1',m_2',m_3',m_4',m_5',m_6',m_7',m_8',m_9') = (-m_1,m_2-n_1,n_1,m_3,m_4,n_1+m_6-n_2,n_2,m_7,m_8)~.
\end{equation}
Imposing $m_i'=\alpha m_i$ $\forall i$ with $\alpha$ an $i$-independent phase yields conditions which also split into orbits:
    \begin{align}
        -m_1 &=e^{i \alpha} m_1~,\\
        (n_1,m_3,m_4,m_5) &=e^{i \alpha}(m_3,m_4,m_5,n_1)~,\\
        (n_2,m_7,m_8,m_9) &=e^{i \alpha}(m_7,m_8,m_9,n_2)~,\\
        m_2-n_1 &= e^{i \alpha} m_2~,\\
        n_1+m_6-n_1 &= e^{i \alpha} m_6.
    \end{align}
If $m_1\neq 0$ then $e^{i \alpha}=-1$, and all remaining masses are determined by, say, $m_1,m_3$ and $m_7$. If $m_1=0$ and at least one of $n_1,m_3,m_4,m_5,n_2,m_7,m_8$ or $m_9$ is non-zero, then $e^{4 i \alpha}=1$ and $e^{i \alpha}\neq 1$. As before, all masses can be expressed in terms of, say, $m_3$ and $m_7$. Last, if only $m_2$ and $m_6$ which connect different orbits are non-zero, then $e^{i \alpha}=1$, and $m_2,m_6$ are free parameters. This generalizes to any configuration of punctures, leading to the count of deformation parameters preserving non-invertible symmetries written in \Cref{Sec:vanishingglobmass}.

The same strategy applies to triality defects of order 6 and 3, as well as to duality and triality symmetries of $\widehat{D}_n$ quivers, with the additional constraint evoked above.

\section{Examples with mass deformations}\label{app:examples}

\subsection{Duality defects}

\subsubsection*{Duality symmetry for \texorpdfstring{$\widehat{D}_4$}{D4} with vanishing global mass}

The configuration for $\widehat{D}_4$ requires 8 marked points to be placed on the torus $\mathbb{T}^2$, and we can organize them in either two orbits of size four, or one orbit of size four and one orbit of size 2, where the latter requires two points to be placed on top of $\mathbb{Z}_2$ fixed points, consequently their $\mathbb{Z}_2$ images sits on the same position. Despite two marked points on the same location would lead to inconsistency in the $\widehat{A}_{n-1}$ case, in the $\widehat{D}_{n}$ case a marked point and its orientifold image can indeed sit on the same point. This is consistent with the construction in \cref{sec:DnUplifts} as well as with the definition of the $\tau_i$, \cref{eq:couplDn}, where none of them vanish in the present configuration. Since the first configuration is discussed in \cref{sec:dualitydefects}, here we examine the second one. To be precise, we place the points as
\begin{align}
    p_1 = \frac{1}{2} \; , \qquad p_3 = i p_2 + 1 \; , \qquad p_4 = \frac{i}{2} \; ,
\end{align}
where $p_2$ is free to be placed with $0 < \mathrm{Re} (p_2) < \mathrm{Im} (p_2) < 1/2$. Starting with this configuration, we can define a non-invertible duality defect as $\mathcal{D} = R_{D,3} \circ \sigma \circ t \circ S$, with $\sigma = s_3 s_2 s_3 s_1 s_2 s_3$ and $t = t_{1}^{(i)} t_{2}^{(i)} t_{3}^{(i)}$. On the masses, we have that
\begin{align}
        \mathcal{D} : (m_0,m_1,m_2,m_3,m_4) \to  (-m_4, -m_3, -m_2, -m_0, -m_1) \, ,
\end{align}
and the solutions of the eigenvalue equation $\mathcal{D} \vec{m} = e^{i \alpha} \vec{m}$ are
\begin{align}
    &\left( m_0 , \, \ldots \, , m_4 \right) = (i, -i, 0, -1, 1 ) m_4 \; , \qquad \mathrm{for} \; \alpha = \frac{\pi}{2} \; , \nonumber \\
    &\left( m_0 , \, \ldots \, , m_4 \right) = (-i, i, 0, -1, 1 ) m_4 \; , \qquad \mathrm{for} \; \alpha = - \frac{\pi}{2} \; , \nonumber \\
    &\left( m_0 , \, \ldots \, , m_4 \right) = (-1, -1, 0, 1, 1 ) m_4 \; , \qquad \mathrm{for} \; \alpha = 0 \; . 
\end{align}
The deformed theory's moduli space is given by
\begin{align}
    x^2 + y^2 w = w \left( w^2 - v^4 \right),  
\end{align}
for both the first and the second solution and with $v=m_4 u$, while the others lead to
\begin{align}
    x^2 + y^2 w = w \left( w + v^2 \right)^2 \; .
\end{align}

\subsection{Triality defects}

We provide some examples with duality symmetries that involve an $ST$ transformation, whose action is discussed around \cref{eq:ST} and that leaves $\tau = \rho = e^{\frac{2 \pi i}{3}}$ invariant. The allowed orbits are the following. Orbits of size one are given be the vertices of the fundamental cell. The single orbit of size two consists of points denoted by $C_1$ and $C_2$ with coordinates,
    \begin{align}
        C_1 = \frac{1}{2} + i \frac{\sqrt{3}}{6} = \frac{\rho+2}{3}\; , \qquad C_2 = i\frac{\sqrt{3}}{3} = \frac{2 \rho + 1}{3} \; ,
    \end{align}
The orbit of size three is realized with the three points
    \begin{align}
        q_1 = \frac{1}{2} \; , \qquad q_2 = \frac{\rho}{2} \; , \qquad q_3 = \frac{\rho+1}{2} \; ,
    \end{align}
while the orbits of size six are given by the points placed at
    \begin{align}
        &p_1 = \alpha \; , \quad p_2 = \rho^2 \alpha - \rho^2 \; , \quad p_3 = - \rho \alpha + \rho \; , \nonumber \\
        &p_4 = 1 + \rho \alpha \; , \quad p_5 = - \rho^2 \alpha \; , \quad p_6 = - \rho^2 - \alpha \; ,
    \end{align}
    with $\alpha$ in the triangle $(0, \, 1, \, \rho +1 )$. 

\subsection*{Mass deformed \texorpdfstring{$\widehat{A}_1$}{A1} quiver theory}

\paragraph{Vanishing global mass}

Consider the theory $\widehat{A}_1$ at $\tau = \rho = e^{i \frac{2 \pi}{3}}$ with an orbit of size 2, and global mass $m = 0$. This configuration preserves a defect $\mathcal{D} = s_1 \circ t_1^{(\rho)} \circ S T$. The unique solution reads
\begin{equation}
    \begin{array}{lll}
       \alpha = - \frac{\pi}{2} \quad & (m_1, - m_1) \quad & xy = w (w - um_1) \; ,
    \end{array}
\end{equation}
which flows to the conifold.

\paragraph{Non-vanishing global mass}

In the case of $\widehat{A}_1$ with global mass $m \neq 0$, the masses transform as
\begin{align}
    - m_1 - m \rho (\tau_1 + 1) & = - \rho m_1 \; , \nonumber \\[4pt]
     m_1 + m \rho \tau_1  & = - \rho m_2 \; , 
\end{align}
whose solution is $\left(\frac{1}{2},1\right) m_2$.

\subsection*{\texorpdfstring{$\widehat{A}_7$}{A7} with vanishing global mass}

\begin{figure}[h!]
    \centering
    \includegraphics[scale=1.2]{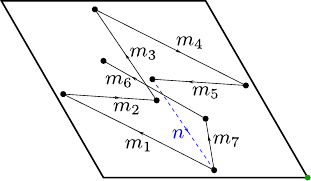}
    \caption{Mass deformations of the $\widehat{A}_7$ quiver.}\label{fig:trialityexample}
\end{figure}

Let us compute the mass deformations of the $\widehat{A}_7$ quiver gauge theory which preserve the non-invertible triality symmetry of order 6. The eight corresponding punctures at $\tau= e^{2i\pi/3}$ necessarily split into an orbit of size 6 and an orbit of size two as displayed in \cref{fig:trialityexample}.

We apply the strategy outline in \cref{app:generalproof} and consider the masses shown in \cref{fig:trialityexample}. Under the action of $ST$ composed with deck transformations and a permutation, the mass deformation preserves the non-invertible triality symmetry if and only if they solve:
\begin{align}
    (n,m_1,m_2,m_3,m_4,m_5) &= e^{i\alpha}(m_1,m_2,m_3,m_4,m_5,n) \nonumber\\
    -m_6 &= e^{i\alpha}m_6 \nonumber \\
    m_1 + m_7 + m_6 &= e^{i\alpha}m_7
\end{align}

We can then read directly that
\begin{itemize}
    \item If $(e^{i\alpha})^6=1$ and $e^{i\alpha}\neq\pm1$, then all masses are determined by $m_1$,
    \item If $e^{i\alpha}=-1$ then all masses are determined by $m_6$ and $m_1$,
    \item If $e^{i\alpha}=1$ then $m_7$ is the only free parameter, as all other masses have to be set to zero.
\end{itemize}
This result translates in any other mass basis, for example the physical one.

\subsection*{Mass deformed \texorpdfstring{$\widehat{D}_4$}{D4} quiver theory}

\paragraph{Vanishing global mass}

Consider the theory $\widehat{D}_4$ at $\tau = \rho = e^{i \frac{2 \pi}{3}}$ with an orbit of size 6 and one orbit of size 2. The configuration preserves a defect $\mathcal{D} = R_{D,1} \circ R_{D,4} \circ s_2 s_1 \circ t_1^{(\rho)}t_2^{(\rho)}t_4^{(\rho)} \circ ST$, which transforms the masses as
\begin{align}
    \mathcal{D} : (m_0, m_1, m_2, m_3, m_4) \to (m_2, m_0 + m_1 + m_2, - m_1 - m_2, m_1 + m_2 + m_4, m_1 + m_2 + m_3) \; .
\end{align}
The solutions are 
\begin{equation}
    \begin{array}{lll}
       \alpha = \pi  \quad & (1,0,-1,0,1)m_0 \quad & x^2 + y^2 w = w^3 + 12 t^2 w^2 + 30 t^4w + 6 t^4 y + 28 t^6 \\[5pt]
       \alpha = \pi  \quad & (1,0,-1,1,0)m_0 \quad & x^2 + y^2 w = w^3 + 12 t^2 w^2 + 30 t^4w + 6 t^4 y + 28 t^6 \\[5pt] 
       \alpha = \frac{\pi}{3}  \quad & (-1,i \sqrt{3},\rho,1,1)m_3 \quad & x^2 + y^2 w = w^3 + ws^4 - w^2 s^2 \\[5pt]
       \alpha = \frac{4\pi}{3}  \quad & (-1,-i \sqrt{3},\rho,1,1)m_3 \quad & x^2 + y^2 w = w^3 + 4 ws^4 - 4 w^2 s^2 - 3 s^6 
    \end{array}
\end{equation}
where $t = u m_0/2$ and $s=u m_3$.

\paragraph{Non-vanishing global mass}
In the case of $\widehat{D}_4$ with global mass $m \neq 0$, the masses transform as 
\begin{align}
    m_2 + m (1 - \tau_1) & = - \rho m_0 \; , \nonumber \\[4pt]
    m_0 + m_1 + m_2 + m \left(\rho^2 - \rho \right)( 1 - \tau_1) & = - \rho m_1 \; , \nonumber \\[4pt]
    - m_1 - m_2 - m \rho^2 (1 - \tau_1) & = - \rho m_2 \; , \nonumber \\[4pt]
    m_1 + m_2 + m_4 + m \left[ - \frac{1}{3} \rho \left( \rho - 1 \right) + \left( \rho + 1 \right) \left( - \rho \tau_1 + \rho - 1 \right) \right] & = - \rho m_3 \; , \nonumber \\[4pt]
    m_1 + m_2 + m_3 + m \left[ \frac{2}{3} \left( \rho^2 - 1 \right) +  \tau_1 \right] & = - \rho m_4 \; , 
\end{align}
whose unique solution is
\begin{align}
    \left( \sqrt{3} \frac{\tau_1 - 1}{\sqrt{3} (3 - \tau_1) + 2 i} \, , \, \frac{3(\tau_1 - 1)(i + 2 \sqrt{3})}{\sqrt{3}(\tau_1 - 3) - 2i} \, , \, \frac{1}{2} \frac{(\tau_1 - 1)(3i + 7 \sqrt{3})}{\sqrt{3}(3 - \tau_1) + 2i} \, , \, 1 + \frac{i + \sqrt{3}}{\sqrt{3}(\tau_1 - 3) - 2i} \, , \, 1 \right) m_4 \; ,
\end{align}
with global mass
\begin{align}
    m = \frac{45 i + 21 \sqrt{3}}{23 \sqrt{3} - 3 (2i + 3 \sqrt{3} )\tau_1 + 36 i} \; .
\end{align}

\bibliographystyle{JHEP}
\bibliography{biblio}

\end{document}